%% file: main.tex
\documentclass{article}

    \usepackage[preprint]{neurips_2022}

\usepackage[utf8]{inputenc} 
\usepackage[T1]{fontenc}    
\usepackage{booktabs}       
\usepackage{amsfonts}       
\usepackage{nicefrac}       
\usepackage{microtype}      
\usepackage[dvipsnames,table]{xcolor}
\usepackage{graphicx}
\usepackage{subfigure}
\usepackage{amsthm}
\usepackage{amssymb}
\usepackage{algorithm}
\usepackage{algorithmic}
\usepackage{multirow}
\usepackage{wrapfig}
\usepackage{enumitem}
\definecolor{citecol}{HTML}{6F130C}
\definecolor{tableofcontent}{HTML}{1F4A83}
\definecolor{urlcol}{HTML}{2470D8}

\usepackage{hyperref}
\hypersetup{
    colorlinks=true,       
    linkcolor=tableofcontent,          
    citecolor=citecol,        
    urlcolor=blue,           
}

\input{math_commands}

\newcounter{bxincomm}
\definecolor{aqua}{rgb}{0.00,0.67,0.80}

\newcounter{ygcounter}

\newcommand{\ygc}[1]{\ygc{\stepcounter{ygcounter}{\bf [YG's comment \arabic{ygcounter}: #1]}\;}}

\usepackage[normalem]{ulem}
\definecolor{blue-violet}{rgb}{0.54, 0.17, 0.89}
\definecolor{mygreen}{rgb}{0.0, 0.5, 0.0}
\definecolor{awesome}{rgb}{1.0, 0.13, 0.32}

\newcommand{\eg}{\textit{e}.\textit{g}.,}
\newcommand{\ie}{\textit{i}.\textit{e}.,}

\title{
How GNNs Facilitate CNNs in Mining Geometric Information from Large-Scale Medical Images
}

\author{%
 Yiqing Shen\textsuperscript{1$*\dag$}, ~ Bingxin Zhou\textsuperscript{2}\thanks{Equal contribution.}, ~ Xinye Xiong\textsuperscript{1}, ~ Ruitian Gao\textsuperscript{1}, ~ Yu Guang Wang\textsuperscript{1,3}\thanks{Corresponding author. Email: \{shenyq, yuguang.wang\}@sjtu.edu.cn.}
 \\[1mm]
 \textsuperscript{1}Shanghai Jiao Tong University, ~ \textsuperscript{2}University of Sydney, ~
 \textsuperscript{3}UNSW\\
}

\begin{document}

\maketitle

\begin{abstract}
Gigapixel medical images provide massive data, both morphological textures and spatial information, to be mined. Due to the large data scale in histology, deep learning methods play an increasingly significant role as feature extractors. Existing solutions heavily rely on convolutional neural networks (CNNs) for global pixel-level analysis, leaving the underlying local geometric structure such as the interaction between cells in the tumor microenvironment unexplored. The topological structure in medical images, as proven to be closely related to tumor evolution, can be well characterized by graphs. To obtain a more comprehensive representation for downstream oncology tasks, we propose a fusion framework for enhancing the global image-level representation captured by CNNs with the geometry of cell-level spatial information learned by graph neural networks (GNN). The fusion layer optimizes an integration between collaborative features of global images and cell graphs. Two fusion strategies have been developed: one with MLP which is simple but turns out efficient through fine-tuning, and the other with \textsc{Transformer} gains a champion in fusing multiple networks. We evaluate our fusion strategies on histology datasets curated from large patient cohorts of colorectal and gastric cancers for three biomarker prediction tasks. Both two models outperform plain CNNs or GNNs, reaching a consistent AUC improvement of more than 5\% on various network backbones. The experimental results yield the necessity for combining image-level morphological features with cell spatial relations in medical image analysis.
Codes are available at \hyperref[https://github.com/yiqings/HEGnnEnhanceCnn]{\color{RoyalBlue}https://github.com/yiqings/HEGnnEnhanceCnn}.
\end{abstract}

\section{Introduction}
\begin{figure*}[tbp!]
    \centering
    \includegraphics[width=\linewidth]{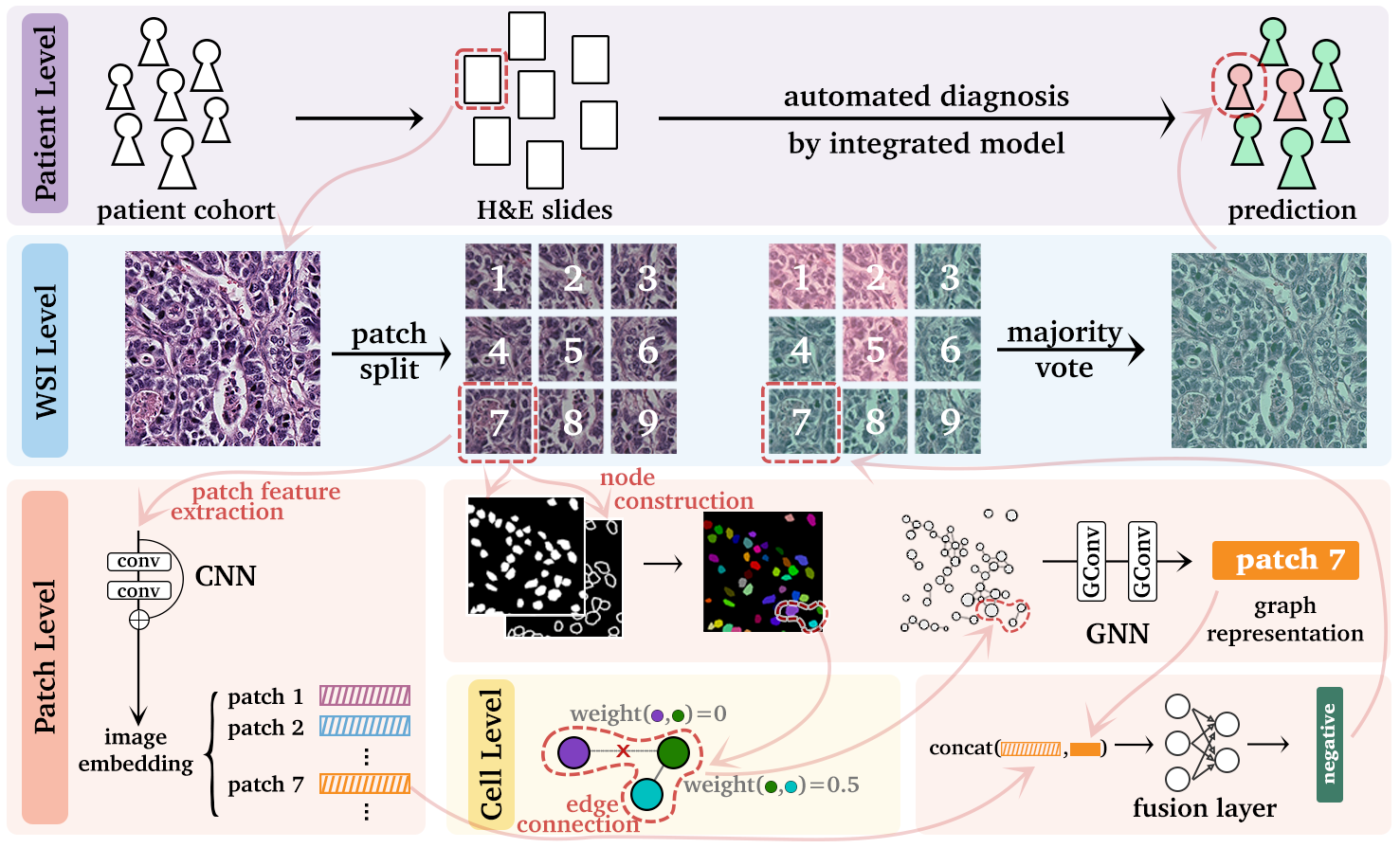}
    \caption{
    The use case of the proposed fusion scheme for GNN and CNN from histology whole-slide image in the patient-level diagnostic task.
    It consists of the following three steps.
    First, for each patches tessellated from WSI, a cell graph is generated to characterize the topological structures, where the nuclei region segmented by CA\textsuperscript{2.5}-Net are identified as graph nodes, with the node features extracted from the pre-defined pathomics. 
    Then, in both the training and inference stage, the global image-level representation together with the geometric representation discovered by CNN and GNN, are integrated by a learnable fusion layer to obtain a more comprehensive feature representation.
    Finally, the patient-level prediction is determined by a majority vote from all patch predictions. 
    }
    \label{fig:architecture_overview}
\end{figure*}

Large-scale medical images, such as histology, provide a wealth of complex patterns for deep learning algorithms to mine.
Existing approaches routinely employ end-to-end convolutional neural networks (CNNs) frameworks, by taking the morphological and textural image features as input. 
Numerous practices with CNNs have been made in diagnostic and prognostic tasks, such as lesion detection, gene mutation identification, molecular biomarker classification, and patient survival analysis from Hematoxylin and Eosin (H\&E) stained histology whole-slide images (WSIs) \citep{shaban2019novel,fu2020pan,liao2020deep,calderaro2021artificial,intro_msi_review}. 
Determined by the convolutional kernel which is primarily targeted to analyze fixed connectivity between local areas (\ie~pixel grids), CNNs focus on extracting global image-level feature representations.
However, no guidance has been imposed explicitly on CNNs to exploit the underlying topological structures from input medical images, \eg~the cell-cell interaction and the spatial distribution of cells, which have been clinically proven to be closely related to tumor evolution and biomarker expression \citep{galon2006type,feichtenbeiner2014critical,barua2018spatial,tumor_spatial_structure}. 
The recognition of the cell dispersal manner and their mutual interactions are essential for training robust and interpretable deep learning models \citep{gunduz2004cell,yener2016cell,wang2021cell}.

Mathematically, the topological structures and cell relationships are formulated by graphs. 
By its definition, a graph can characterize the relationship between nodes, \eg~super-pixels in natural images, or the cells in histological images. 
Following the establishment of graphs, graph neural networks (GNNs) were proposed to learn the geometric information \citep{bronstein2017geometric,wu2020comprehensive,zhang2020deep}. 
While CNNs are capable of learning global image representation, GNNs can provide machinery for the local topological features. 
Both global and local features serve as significant representation in learning the mapping of histological image space to clinical meaningful biomarkers.
One strategy is to make use of the own merits of CNN and GNN models. 
Some recent attempts at combining GNNs with CNNs have achieved satisfactory performance boost in natural image classification tasks, such as remote sensing scene recognition \citep{liang2020deep, peng2022multi} and hyper-spectral image prediction \citep{dong2022weighted}. 
In the medical imaging domain, \cite{wei2022collaborative} predicted isocitrate dehydrogenase gene mutation with a collaborative learning framework that aligns a CNN for tumor MRI with a GNN for tumor geometric shape analysis. 
To the best of our knowledge, a study of the interplay between CNNs and GNNs for histology is still absent. 

In this paper, we develop an efficient strategy that is able to integrate the structure feature from GNNs with the image feature of CNNs for H\&E slides analysis.
The fusion scheme partitions a WSI into non-overlapped patches and generates a cell graph for each patch by linking associated cells (see Section~\ref{sec:method}) to model the cell interactions. 
Then, a GNN is employed to distill geometric representation. 
To fuse the graph-level representation learning with image-level embedding, we train the GNN together with the CNN in parallel.
The integration takes place in a learnable fusion layer which incorporates the morphology feature of the whole image with the geometric representation of cell graphs. In this way, insights into the spatial structure are gained for a specific staining image, such as the distribution of cells, interaction of cancer and healthy cells, and tumor microenvironment.

In practice, we can simply connect a learnable fusion layer using MLP or \textsc{Transformer} next to the outputs of GNN and CNN modules. The simple amalgamation can produce a model which outperforms a sole GNN or CNN model on real histological image datasets (two public and one private).
The key to performance improvement of the fusion model lies in that the local geometry of the cell graphs of patches which can only be perceived by GNNs tops up the global image feature of CNNs.

\paragraph{Contributions.} The contributions are three-fold: 
(1) We develop two fusion schemes, based on MLP and \textsc{Transformer}, for integrating the features extracted from CNNs and GNNs.
Moreover, we present a use case of the proposed framework on histology analysis, where cell geometric behaviors are crucial for downstream diagnostics. 
(2) Experiments on three real-world and one synthetic datasets yield that geometric and image-level representations are complementary.
(3) We release the constructed graph-image paired datasets, which can serve as a benchmark for future research in the image-graph bimodal domain.

\section{Fusing CNN with GNN}
\label{sec:method}

\paragraph{Problem Formulation.}
CNNs extract the global image-level representation $\mH_{\gI}$ from an input image $\mX_\gI$. 
However, the underlying geometric relationship, characterized by a graph $(\mX_\gG,\mA_\gG)$ with node feature $\mX_\gG$ and adjacent matrix $\mA_\gG$, is not explicitly explored in CNN, although it is crucial for tasks such as medical imaging analysis. 
Therefore, we leverage a GNN to capture the geometric representation $\mH_{\gG}$ as an enhancement to $\mH_{\gI}$. 
The major scope of this paper is to construct a fusion scheme for the bimodal data $\big(\mX_\gI,(\mX_\gG,\mA_\gG)\big)$, especially in the medical domain.

\subsection{Geometric Feature Representation} 
We denote the corresponded graph (\eg~cell graph in histology) to the image $\mX_\gI$ as $\gG=(\gV,\gE)$, where $\gV$ is the collection of nodes, $\gE$ is the set of all edges $e_{ij}$ with the attribute $w_{ij}$ describing the pair-wise node interaction. For notation simplicity, we use a matrix pair $(\mX_\gG,\mA_\gG)$ to represents the node attributes and weighted edges, respectively. We name $\mA_\gG$ as an adjacency matrix with its element $(\mA_\gG)_{ij}=w_{ij}$. 
The $\ell$th layer of the GNN finds the hidden representation of the graph by
\begin{equation} \label{eq:gnn}
    \mH^{\ell}_{\gG}=
    \texttt{ReLU}
    \big(\texttt{GraphConv}\left(\mA_\gG,\mH^{\ell-1}_\gG\right)\big),
\end{equation}
where $\mH^{0}_\gG=\mX_\gG$. 
We consider spatial-based convolutions for $\texttt{GraphConv}$, which usually follow the message-passing \citep{gilmer2017neural} form.
For the $i$th node of a graph, its representation $\mH_{\gG,i}^{\ell}$ at the $\ell$th convolutional layer reads
\begin{equation} \label{eq:mpnn}
    \mH_{\gG,i}^{\ell} = \gamma\left(\mH_{\gG,i}^{\ell-1}, \square_{j\in \mathcal{N}(i)} \phi(\mH^{\ell-1}_{\gG,i},\mH^{\ell-1}_{\gG,j},\mA_{ij})\right)
\end{equation}
with some differentiable operators $\gamma$, $\phi$ (\eg~MLP) and permutation invariant aggregation function $\square$ (\eg~average or summation). The $\mH^{\ell-1}_{\gG,j}$ denotes node $j$'s hidden representation at the $(\ell-1)$th layer, where $j\in\gN(i)$ is a 1-hop neighbor of node $i$, \ie~ $\mA_{ij}=\mA_{ji}\neq0$.

The representation $\mH^{\ell}_{\gG}$ embeds spatial topological structures of the underlying graph, which is usually sent to a readout layer, such as a linear layer, before eventually being fed into the fusion layer. 
We term this linear layer as the \textit{alignment layer}, which helps to align the feature dimensions of the GNN with the parallel CNN output.

\begin{figure}[tbp!]
    \centering
    \includegraphics[width=\linewidth]{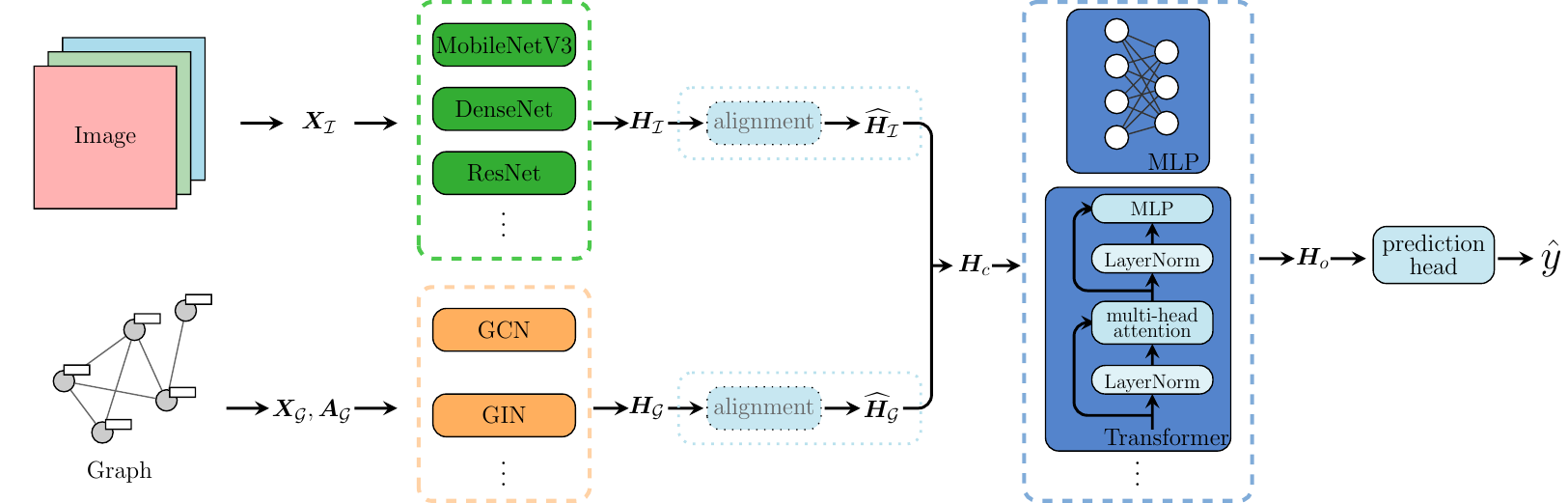} 
    \caption{
    A schematic illustration for fusing the representations of local geometry and global image features. 
    An image input $\mX_\gI$ is encoded to a global image-level presentation $\mH_\gI$ by CNNs. 
    Meanwhile, the geometric information $(\mX_\gG,\mA_\gG)$ from the image, which is first transformed into a cell graph with attributes, is embedded by GNNs (\ie~$\mH_\gG$). 
    The two sets of hidden representation are then fed into a fusion layer for adaptive integration (\ie~$\mH_c$). 
    The output $\mH_o$ is eventually send to a prediction head (\eg~classification or regression head) for training assignment.}
    \label{fig:architecture_model}
\end{figure}

\subsection{Image-level Feature Representation}
The image-level feature representation is directly extracted from histology patches by CNNs.
For instance, denote $\{\mH^{1}_{\gI},\dots,\mH^{\ell-1}_{\gI}\}$ the output of the first $(\ell-1)$ blocks after convolution layers. 
We can use different convolutional module for the CNN. For example, a \textsc{DenseNet} \citep{huang2017densely} defines
\begin{equation}\label{eq:DenseNet}
    \mH^{\ell}_{\gI}=
    \texttt{ReLU}
    \big(\texttt{Conv}(\texttt{concat}[\mH^{1}_{\gI},\dots,\mH^{\ell-1}_{\gI}])\big).
\end{equation}
Alternatively, \textsc{ResNet} \citep{he2016deep} finds $\mH^{\ell}_{\gI}$ by
\begin{equation}\label{eq:ResNet}
    \mH^{\ell}_{\gI} = \texttt{ReLU}\big(\texttt{Conv}(\mH^{\ell-1})+\mH^{\ell-1}\big)
\end{equation}
with some activated convolutional layers $\texttt{Conv}(\cdot)$. 
The residual connection in the second design can reduce the computational cost of deep CNNs and circumvent gradient diminishing. 
In the empirical study, a lightweight architecture namely \textsc{MobileNetV3} \citep{howard2019searching} is considered, where efficient depth-wise separable convolutions along with the inverted residuals replace traditional convolution layers. 
%
For all CNN blocks, we assign the input feature $\mH^{0}_{\gI}$ by staining normalized histology image patches. 
In the same fashion as geometric feature representation, the final image representation is fed to a learnable fully-connected layer to adjust the embedding feature dimensions.

\subsection{Learnable Feature Fusion Layer}
Denote the output representations from image and graph as $\mH_{\gI}\in\mathbb{R}^{d_\gI}$ and $\mH_{\gG}\in\mathbb{R}^{d_\gG}$.
We then train the fusion layer to learn the optimal integration between them. 
In particular, we consider two candidate structures: MLP and \textsc{Transformer} \citep{vaswani2017attention} for feature fusion. 
The former approaches the fused representation $\mH_o$ with \texttt{MLPBlock}s formulated as 
\begin{equation}
    \mH_o  =
        \texttt{MLPBlock}(...( \texttt{MLPBlock}(\mH_c))),
        \label{eq:mlp}
\end{equation}
where $\mH_c = \texttt{concat}\big[\mH_{\gI} ,\mH_{\gG}\big]\in \mathbb{R}^{d_\gI+d_\gG}$, and 
\begin{equation*}
    \texttt{MLPBlock}(\mH) = \texttt{Dropout}(\texttt{ReLU}(\texttt{Linear}(\mH))).
\end{equation*}
The \textsc{Transformer} fusion scheme formulates $\mH_o$ by
\begin{equation}
\mH_o = \texttt{Pooling}(
        \texttt{TransBlock}(\cdots( \texttt{TransBlock}(\mH_c)))), \label{eq:transformer}
\end{equation}
where $\mH_c = \texttt{stack}\big[\mH_{\gI}, \mH_{\gG}\big] \in \mathbb{R}^{2\times d} $. 
The \texttt{stack} operation requires an identical dimension of $\mH_{\gI}$ and $\mH_{\gG}$, thus feature shape alignment with additional linear layer is required \ie~ $d=d_\gI=d_\gG$.
The \texttt{TransBlock} represents the PreNorm variant of Transformer \citep{wang2019learning}, \ie
\begin{equation*}
    \texttt{TransBlock}(\mH) =\texttt{ResidualPreNorm}(\texttt{MLPBlock}, 
    \texttt{ResidualPreNorm}(\texttt{MHSA},\mH)),
\end{equation*}
where \texttt{MHSA} is the multi-headed self attention layer, and we use \texttt{LayerNorm} as the normalization layer.

\paragraph{Extension to Multiple Networks.}
When multiple numbers of CNNs and/or GNNs are involved, we write the extracted representations from CNNs and GNNs as $\mH_\gI^i \in \mathbb{R}^{d_\gI^i}$ (for $i=1,\cdots,k_\gI$), $\mH_\gG^j \in \mathbb{R}^{d_\gG^j}$ (for $j=1,\cdots,k_\gG$), with $d_\gI^i$ and $k_\gG$ denote the number of CNN and GNN respectively.
%
We write the representations after feature shape alignment (\ie~a linear layer) as $\widehat{\mH}_\gI^i \in \mathbb{R}^{d}$ and $\widehat{\mH}_\gG^j \in \mathbb{R}^{d}$. 
In the MLP fusion scheme by Eq. \eqref{eq:mlp}, we formulate $\mH_c$ by 
\begin{equation}
    \mH_c = \texttt{concat}\big[{\mH}_\gI^1,\cdots,{\mH}_\gI^{k_\gI}, \mH_\gG^1,\cdots,\mH_\gG^{k_\gG}\big]\in \mathbb{R}^{\sum d_\gI^i+\sum d_\gG^j}.
\end{equation}
In the \textsc{Transformer} fusion scheme by Eq. \eqref{eq:transformer}, $\mH_c$ turns to be 
\begin{equation}
    \mH_c = \texttt{stack}\big[\widehat{\mH}_\gI^1,\cdots,\widehat{\mH}_\gI^{k_\gI}, \widehat{\mH}_\gG^1,\cdots,\widehat{\mH}_\gG^{k_\gG}\big]\in \mathbb{R}^{(k_\gI+k_\gG)\times d}.
\end{equation}

\paragraph{Feature Alignment Layer.} 
The feature shape alignment layer is a learnable linear layer, transferring the hidden representations to fixed output shape \ie
\begin{equation}
    \widehat{\mH}= \texttt{Linear}(\mH) \in \mathbb{R}^d.
\end{equation}
We provide three alignment strategies, depending on the aligned feature shape (\ie~ value of $d$) 
\begin{itemize}[leftmargin=*]
    \item \textit{Minimization Alignment}: Set the output size as $d = \min \{d_\gI^i, d_\gG^j\}$, which can reduce the size of $\mH_c$. Thus the computational cost can be alleviated, especially for the multi-headed self attention layers in \textsc{Transformer}. 
    \item \textit{Maximization Alignment}: Set $d = \max \{d_\gI^i, d_\gG^j\}$. This strategy aims not to compress the features, thus brings intensive computations. 
    \item \textit{Pre-defined Shape Alignment}: Use a manually assigned $d$. 
\end{itemize}

\paragraph{Prediction Head.} 
The final prediction for classification or regression tasks, based on the fused representation is 
\begin{equation}
    \hat{y} = \texttt{Linear}(\mH_o).
\end{equation}

\section{Fusion Model For Medical Image Analysis}

\begin{figure}[t]
    \centering
    \begin{minipage}{0.23\textwidth}
        \includegraphics[width=\linewidth]{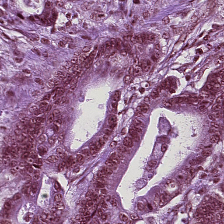}
    \end{minipage}
    \hspace{1mm}
    \begin{minipage}{0.23\textwidth}
        \includegraphics[width=\linewidth]{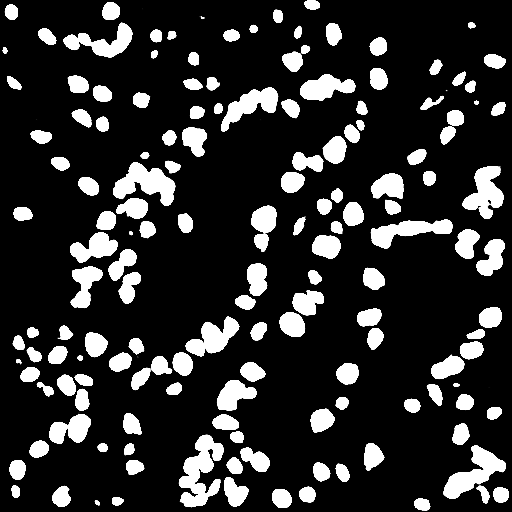}
    \end{minipage}
    \hspace{1mm}
    \begin{minipage}{0.23\textwidth}
        \includegraphics[width=\linewidth]{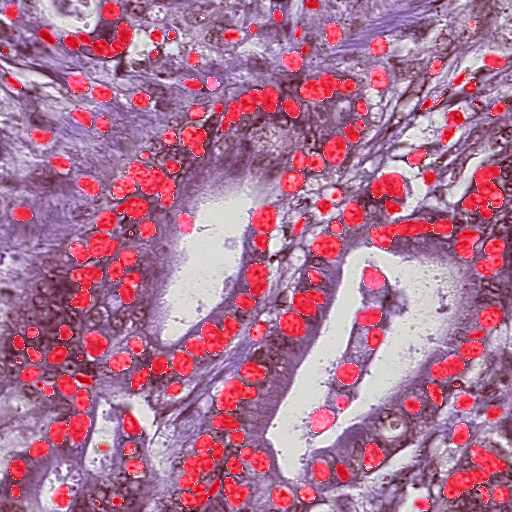}
    \end{minipage}
    \hspace{1mm}
    \begin{minipage}{0.23\textwidth}
        \includegraphics[width=\linewidth]{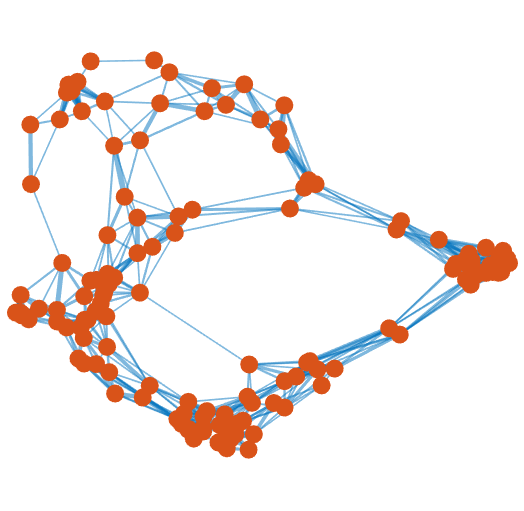}
    \end{minipage}
    \caption{Visualization of the segmented cells and the generated graphs from an arbitrary patch sample of \textbf{GIST-PDL1}. The four subgraphs from left to right are the raw patch image, the segmented cells masks, the patch image with overlaid segmentation masks, and the generated graph. }
    \label{fig:nucleiSeg_graph_PDL1}
\end{figure}

\paragraph{Cell Graph.}
We establish a cell graph for each patch image, which process is visualized in Figure~\ref{fig:nucleiSeg_graph_PDL1}.
The graph nodes ($v_i$, with its subscript $i$ representing the node index) in a cell graph are biologically determined by the nuclei regions. 
The cell graph can represent the cell-cell interaction and the collection of cell graphs for all patches provide a precise characterization of tumor microenvironment. 
With only the availability of raw image patches, we leverage the nuclei regions segmented by a well-tuned CA\textsuperscript{2.5}-Net \citep{segment} to extract the node features of each single nuclei node (See Appendix~\ref{sec:app:seg}). 
We follow \cite{lambin2017radiomics} and extract a total number of $94$ pre-defined pathomics features $\mX_\gG$ for each nuclei region as the corresponding graph node feature. The details are elaborated in Appendix~\ref{sec:app:tmeAttribute}.
As the morphological signals are believed relative to cell-cell interplay, the cell-specific features $\mX_\gG$, which include the nuclei coordination, optical, and representations, then characterize the cell-level morphological behavior.

We then calculate the pair-wise Euclidean distance between nuclei centroids to establish edges of a cell graph \citep{wang2021cell} to quantify the interplay between cells in a patch.
To be precise, for arbitrary two nuclei nodes $v_i$ and $v_j$, with their associated centroid Cartesian coordinates $(x_i,y_i)$ and $(x_j,y_j)$, the edge weight $w_{ij}$ for the interaction between two nodes reads
\begin{equation}
w_{ij}:= 
\begin{cases}
\displaystyle{d_c}/{d(v_i,v_j)},~&d(v_i,v_j)\leq d_c~\text{pixels},\\[2mm]
0,~&\text{otherwise},\\
\end{cases}
\end{equation}
where $d(v_i,v_j)$ regards the Euclidean distance between $v_i$ and $v_j$. 
From the clinical observations, two cells do not exert mutual influence with their centroid distance exceeding $d_c$ \citep{barua2018spatial}.
Thus, the critical distance $d_c$ depicts the range where a cell can interact with another.
Note that the precise value of $d_c$ depends on the tissue structure, image category, and magnification of the WSI. 
An edge $e_{ij}$ exists between $v_i$ and $v_j$ if and only if the weight $w_{ij}>0$.
When a patch is acquired as a cell graph $\gG$, its geometric feature representation can be gradually learned by a GNN. 

\paragraph{Overall Pipeline with the Fusion Model.}
%
%
A use case of the proposed GNN and CNN fusion scheme for the downstream patient-level prediction from histology is presented in Figure.~\ref{fig:architecture_overview}.
First, we partition a WSI into non-overlapped patches of the same size, \eg~ $224\times224$ pixels in this research. 
Subsequently, a cell graph is extracted for each patch to characterize the topological structure of the local cell behavior, where the nodes are identified as the nuclei region segmented by CA\textsuperscript{2.5}-Net \citep{segment}. 
In the training stage, GNN and CNN simultaneously extract the geometric representation from the cell graph and the global image-level presentation.
Finally, the output image and graph embeddings are fused by a learnable layer with MLP or \textsc{Transformer} (Figure~\ref{fig:architecture_model}).

\section{Experiments}
\label{sec:exp}

\subsection{Implementation and Setups}


\paragraph{Dataset.} We evaluate the proposed fusion scheme on three real-world H\&E stained histology benchmarks, termed as \textbf{CRC-MSI}, \textbf{STAD-MSI}, and \textbf{GIST-PDL1}. 
The first two datasets targets binary microsatellite instability (MSI) status classification \citep{msi0}, where we follow the original train and test split. 
%
We also evaluate the performance of the model on a binary Programmed Death-Ligand 1 (PD-L1) status binary classification dataset, which was curated from $129$ well-annotated WSIs of gastric cancer patients. 
We supplement further details for data collection and descriptions in Table \ref{table:descriptive} and Appendix~\ref{sec:app:data}.

\begin{table}[th]
\caption{Summary of the three datasets.}
\label{table:descriptive}
\begin{center}
\begin{tabular}{llrrr}
\toprule
&\textbf{Dataset} & \textbf{GIST-PDL1} & \textbf{CRC-MSI} & \textbf{STAD-MSI} \\ 
\midrule
\multirow{7}{*}{\rotatebox[origin=c]{90}{IMAGE}}\hspace{0.1cm} 
& \# Patients            & $129$     & $315$     & $360$ \\
& \# Training Images     & $7,676$   & $93,408$  & $100,570$ \\
& Training Positive Rate & $41.10 \%$& $50.0\%$  & $50.0\%$  \\
& \# Test Images         & $2,471$   & $99,904$  & $118,008$ \\
& Test Positive Rate     & $47.71\%$ & $29.4\%$  & $23.6\%$  \\
& Magnification          & $20\times$& $20\times$& $20\times$ \\ 
& Original Patch Size    & $512\times512$ &  $224\times224$ & $224\times224$  \\
\midrule
\multirow{5}{*}{\rotatebox[origin=c]{90}{GRAPH}}\hspace{0.1cm} 
& Min \# Nodes           & $50$      &  $1$      &  $1$ \\
& Max \# Nodes           & $621$     &  $103$    &  $120$ \\
& Median \# Nodes        & $199$     &  $40$     &  $51$ \\
& Avg \# Nodes           & $206$     &  $40$     &  $50$ \\
& Avg \# Edges           & $3,402$   &  $163$    &  $246$ \\
\bottomrule
\end{tabular}
\end{center}
\end{table}

\paragraph{Model Configurations.}
We evaluate the performance gain of our proposed fusion scheme with a comprehensive comparison against three CNN backbones of different scales:
\textsc{MobileNetV3} \citep{howard2019searching}, \textsc{DenseNet} \citep{huang2017densely}, and \textsc{ResNet} \citep{he2016deep}. 
We stack two graph convolution layers for graph representation learning. Two candidates of graph convolution \textsc{GCN} \citep{kipf2017semi} and \textsc{GIN} \citep{xu2018powerful} are taken into account, following a $2$-layer \textsc{TopK} \citep{cangea2018towards} graph pooling scheme. The graph convolution plays the critical role in extracting the geometric feature of the patch.
For the fusion layer, both a $1$-layer MLP and \textsc{Transformer} are validated.
We name the models in Table~\ref{tab:classification} with the adopted model architectures and modules. For instance, \textsc{ResNet}-\textsc{GCN}-\textsc{MLP} indicates a \textsc{ResNet} for image embedding, \textsc{GCN} plus \textsc{TopK} for graph representation learning, and \textsc{MLP} \texttt{MLPBlock} for features fusion. We fix the number of layers for each block. For example, we take $121$ convolution layers for \textsc{DenseNet} in all five related models.
Details of the model configurations and training hyper-parameters are elaborated in Appendix~\ref{sec:app:expDetail}.

\begin{table*}[t]
    \caption{Average test ACC and AUC comparisons on three benchmarks over $7$ repetitions.}
    \label{tab:classification}
    \begin{center}
    \resizebox{\linewidth}{!}{
    \begin{tabular}{llrrrrrrrrr}
    \toprule
    && \multicolumn{3}{c}{\textbf{GIST-PDL1}} & \multicolumn{3}{c}{\textbf{CRC-MSI}} & \multicolumn{3}{c}{\textbf{STAD-MSI}}\\ \cmidrule(lr){3-5}\cmidrule(lr){6-8}\cmidrule(lr){9-11}
    \multicolumn{2}{c}{\textbf{Model}} & ACC & AUC & AUC\textit{\footnotesize{patient}} & ACC & AUC & AUC\textit{\footnotesize{patient}} & ACC & AUC & AUC\textit{\footnotesize{patient}} \\
    \midrule
    &\textsc{GCN} & $68.5${\scriptsize $\pm1.71$} & $73.3${\scriptsize $\pm1.33$} & $58.2${\scriptsize $\pm2.54$} & $66.8${\scriptsize $\pm2.12$} & $56.5${\scriptsize $\pm1.48$} & $51.1${\scriptsize $\pm6.52$} & $69.3${\scriptsize $\pm6.75$} & $55.9${\scriptsize $\pm1.31$} & $62.9${\scriptsize $\pm1.91$} \\
    &\textsc{GIN} & $71.9${\scriptsize $\pm1.37$} & $77.2${\scriptsize $\pm0.81$} & $62.9${\scriptsize $\pm1.43$} & $66.6${\scriptsize $\pm1.79$} & $57.0${\scriptsize $\pm0.91$} & $44.6${\scriptsize $\pm3.39$} & $71.3${\scriptsize $\pm3.52$} & $60.3${\scriptsize $\pm1.22$} & $66.8${\scriptsize $\pm4.09$}\\
    \midrule 
    \multirow{5}{*}{\rotatebox[origin=c]{90}{MobileNetV3}}
    & NA & $73.6${\scriptsize $\pm2.14$} & $86.8${\scriptsize $\pm1.24$} & $83.9${\scriptsize $\pm4.08$} & $73.0${\scriptsize $\pm0.75$} & $66.3${\scriptsize $\pm0.55$} & $64.7${\scriptsize $\pm3.09$}
    & $75.3${\scriptsize $\pm1.26$} & $66.9${\scriptsize $\pm2.11$} & $72.8${\scriptsize $\pm1.15$} \\
    &\textsc{GCN}-\textsc{MLP} & $77.2${\scriptsize $\pm0.68$} & $88.7${\scriptsize $\pm0.51$} & $92.0${\scriptsize $\pm0.36$} & $72.7${\scriptsize $\pm0.34$} & \bm{$73.5${\scriptsize $\pm0.60$}} & $77.5${\scriptsize $\pm2.82$} & $76.1${\scriptsize $\pm0.42$} & $71.9${\scriptsize $\pm0.86$} & $73.1${\scriptsize $\pm0.99$}\\
    &\textsc{GIN}-\textsc{MLP} & $74.9${\scriptsize $\pm1.23$} & $89.4${\scriptsize $\pm0.36$} & $94.2${\scriptsize $\pm2.28$} & $73.4${\scriptsize $\pm0.20$} & $69.1${\scriptsize $\pm3.80$} & $78.1${\scriptsize $\pm3.98$} & $75.8${\scriptsize $\pm0.37$} & $69.8${\scriptsize $\pm1.01$} & $73.2${\scriptsize $\pm0.79$}\\
    &\textsc{GCN}-Trans & \bm{$77.9${\scriptsize $\pm1.17$}} & $90.5${\scriptsize $\pm0.86$} & \bm{$96.7${\scriptsize $\pm1.02$}} & $73.2${\scriptsize $\pm0.44$} & $71.0${\scriptsize $\pm0.76$} & $77.8${\scriptsize $\pm3.70$} & \bm{$76.3${\scriptsize $\pm0.41$}} & \bm{$73.6${\scriptsize $\pm0.70$}} & $74.3${\scriptsize $\pm0.64$} \\
    &\textsc{GIN}-Trans & $76.2${\scriptsize $\pm1.37$} & \bm{$90.9${\scriptsize $\pm0.86$}} & $94.4${\scriptsize $\pm1.80$} & \bm{$73.5${\scriptsize $\pm0.37$}} & $70.5${\scriptsize $\pm1.33$} & \bm{$79.5${\scriptsize $\pm2.40$}} & $76.3${\scriptsize $\pm0.68$} & $73.4${\scriptsize $\pm1.44$} & \bm{$74.4${\scriptsize $\pm1.51$}}\\
    \midrule
    \multirow{5}{*}{\rotatebox[origin=c]{90}{DenseNet}}
    & NA & $71.2${\scriptsize $\pm1.42$} & $82.5${\scriptsize $\pm3.25$} & $89.0${\scriptsize $\pm4.29$} & $74.2${\scriptsize $\pm0.28$} & $70.0${\scriptsize $\pm0.91$} & $66.7${\scriptsize $\pm4.83$} & $74.9${\scriptsize $\pm1.68$} & $65.5${\scriptsize $\pm1.08$} & $74.9${\scriptsize $\pm0.03$} \\
    &\textsc{GCN}-\textsc{MLP} & $76.5${\scriptsize $\pm0.90$} & $88.9${\scriptsize $\pm0.78$} & $95.9${\scriptsize $\pm3.50$} & \bm{$75.2${\scriptsize $\pm0.29$}} & $70.0${\scriptsize $\pm1.08$} & $68.2${\scriptsize $\pm2.22$} & $76.6${\scriptsize $\pm0.40$} & $74.5${\scriptsize $\pm0.99$} & $75.9${\scriptsize $\pm1.24$}\\
    &\textsc{GIN}-\textsc{MLP} & $76.7${\scriptsize $\pm0.71$} & $88.0${\scriptsize $\pm0.67$} & $94.6${\scriptsize $\pm1.19$} & $74.6${\scriptsize $\pm0.24$} & $70.1${\scriptsize $\pm0.88$} & $71.6${\scriptsize $\pm1.32$} & \bm{$77.0${\scriptsize $\pm0.33$}} & \bm{$74.8${\scriptsize $\pm1.13$}} & $75.4${\scriptsize $\pm0.74$}\\
    &\textsc{GCN}-Trans & \bm{$79.6${\scriptsize $\pm0.76$}} & \bm{$89.8${\scriptsize $\pm1.08$}} & \bm{$97.5${\scriptsize $\pm1.57$}} & $74.9${\scriptsize $\pm0.36$} & $74.0${\scriptsize $\pm0.59$} & \bm{$83.2${\scriptsize $\pm2.05$}} & $76.7${\scriptsize $\pm0.79$} & $73.6${\scriptsize $\pm0.71$} & \bm{$76.3${\scriptsize $\pm1.06$}}  \\
    &\textsc{GIN}-Trans & $75.8${\scriptsize $\pm1.51$} & $87.7${\scriptsize $\pm1.09$} & $96.0${\scriptsize $\pm2.00$} & $74.8${\scriptsize $\pm0.54$} & \bm{$74.6${\scriptsize $\pm0.63$}} & $75.7${\scriptsize $\pm3.17$} & $76.8${\scriptsize $\pm0.58$} & $74.4${\scriptsize $\pm1.00$} & $75.8${\scriptsize $\pm0.95$}\\
    %
    \midrule
    \multirow{5}{*}{\rotatebox[origin=c]{90}{ResNet}}
    & NA & $70.7${\scriptsize $\pm2.09$} & $82.1${\scriptsize $\pm1.53$} & $86.3${\scriptsize $\pm1.65$} & $73.5${\scriptsize $\pm0.40$} & $65.3${\scriptsize $\pm3.95$} & $61.7${\scriptsize $\pm4.97$} & $73.8${\scriptsize $\pm1.51$} & $72.6${\scriptsize $\pm0.95$} & $74.1${\scriptsize $\pm0.82$} \\
    &\textsc{GCN}-\textsc{MLP} & \bm{$81.9${\scriptsize $\pm3.46$}} & \bm{$92.6${\scriptsize $\pm1.52$}} & $94.1${\scriptsize $\pm3.43$} & $74.2${\scriptsize $\pm0.43$} & \bm{$75.2${\scriptsize $\pm0.85$}} & $83.2${\scriptsize $\pm1.26$} & $76.0${\scriptsize $\pm0.63$} & $73.1${\scriptsize $\pm0.79$} & $74.8${\scriptsize $\pm0.75$} \\
    &\textsc{GIN}-\textsc{MLP} & $76.3${\scriptsize $\pm2.02$} & $87.6${\scriptsize $\pm2.48$} & $91.3${\scriptsize $\pm2.57$} & $74.5${\scriptsize $\pm0.70$} & $69.9${\scriptsize $\pm2.19$} & $83.0${\scriptsize $\pm1.67$} & \bm{$76.2${\scriptsize $\pm0.42$}} & \bm{$74.7${\scriptsize $\pm0.89$}} & $75.6${\scriptsize $\pm2.17$} \\
    &\textsc{GCN}-Trans & $76.0${\scriptsize $\pm1.78$} & $86.4${\scriptsize $\pm2.68$} & $93.7${\scriptsize $\pm3.64$} & \bm{$74.8${\scriptsize $\pm0.32$}} & $73.6${\scriptsize $\pm1.28$} & \bm{$84.7${\scriptsize $\pm2.11$}} & $76.1${\scriptsize $\pm0.80$} & $72.7${\scriptsize $\pm0.45$} & $75.6${\scriptsize $\pm1.23$} \\
    &\textsc{GIN}-Trans & $76.8${\scriptsize $\pm1.08$} & $92.1${\scriptsize $\pm0.57$} & \bm{$95.5${\scriptsize $\pm0.94$}} & $74.6${\scriptsize $\pm0.47$} & $73.4${\scriptsize $\pm0.77$} & $83.8${\scriptsize $\pm1.65$} & $76.2${\scriptsize $\pm0.41$} & $74.6${\scriptsize $\pm0.24$} &  \bm{$75.8${\scriptsize $\pm1.71$}}\\
    \bottomrule\\[-2.5mm]
    \end{tabular}
    }
    \end{center}
\end{table*}

\subsection{Results and Analysis}
Image-level performance is evaluated with two metrics, namely test accuracy (ACC) and area-under-curve (AUC). 
Similarly, we evaluate patient-level prediction with AUC (denoted as AUC\textit{\footnotesize{patient}}). 
As shown in Table~\ref{tab:classification}, the fused learning schemes achieve more than $5\%$ performance gain over plain CNNs. The improvement is more significant at the patient level at up to $23\%$. The additional performance boost suggests that our design of the integrated scheme has better potential to overcome the disturbance of heterogeneous patches for patient-level overall diagnosis. 
The main takeaways include:
1) An individual GNN fails to achieve satisfactory performance. But as a parallel layer, GNNs can enhance the learning capability of CNN by a learnable fusion layer. 2) \textsc{MLP}, though simple, serves as a good fusion layer. 3) Generally speaking, the \textsc{Transformer} integrator outperforms the simple MLP scheme. However, one can not tell whether MLP or \textsc{Transformer} is a universally better fusion solution.
4) All the integrated models outperform the plain CNNs or GNNs. 
5) For the choice of a GNN module, GCN and GIN do not present a significant advantage one over the other.
More empirical investigations are supplemented in Appendix~\ref{sec:app:analysis}, including training cost, performance improvement rate, as well as the performance ranking.

We also investigate the overall ranking of each model fusion configuration in Table~\ref{tab:ranking}. 
We report the averaged ranking for every CNN architecture over three datasets, where in each dataset ranks are calculated by sorting the reported scores.
Generally, the \textsc{Transformer} integrator outperforms the simple MLP scheme. 
For the choice of GNNs, GCN and GIN do not present a significant advantage one over another. 
Nevertheless, all the integrated models outperform designs with plain CNNs. 

\begin{table}[th]    
    \caption{Performance ranking reports the average
rank across three datasets. For simplicity, \textsc{MV3},\textsc{Dense} and \textsc{Res} denote \textsc{MobileNetV3}, \textsc{DenseNet-121} and \textsc{ResNet-18} respectively. The `\textbf{Avg}' columns report the averaged ranking over three CNN architectures.}
    \label{tab:ranking}
    \begin{center}
    \resizebox{\linewidth}{!}{
    \begin{tabular}{lccccccccccccc}
    \toprule
     & \multicolumn{4}{c}{ACC} & \multicolumn{4}{c}{AUC} & \multicolumn{4}{c}{AUC\textit{\footnotesize{patient}}} & \\ \cmidrule(lr){2-5} \cmidrule(lr){6-9} \cmidrule(lr){10-13} 
     & \textsc{MV3} & \textsc{Dense} & \textsc{Res} & \textbf{Avg} & \textsc{MV3} & \textsc{Dense} & \textsc{Res} & \textbf{Avg} & \textsc{MV3} & \textsc{Dense} & \textsc{Res} & \textbf{Avg}& \textbf{Overall} \\
     \midrule
    CNN & 4.7 & 5.0 & 5.0 & \cellcolor{black!10!}4.9 & 5.0 & 5.0 & 5.0 & \cellcolor{black!10!}5.0 & 5.0 & 5.0 & 5.0 & \cellcolor{black!10!}5.0 & \cellcolor{black!20!}5.0 \\
    GCN-MLP & 3.3 & 2.7 & 3.0 & \cellcolor{black!10!}3.0 & 2.3 & 2.7 & 1.7 & \cellcolor{black!10!}2.2 & 4.0 & 3.0 & 3.0 & \cellcolor{black!10!}3.3 & \cellcolor{black!20!}2.9 \\
    GCN-\textsc{Trans} & 3.3 & 2.3 & 2.3 & \cellcolor{black!10!}2.7 & 4.0 & 2.3 & 2.7 & \cellcolor{black!10!}3.0 & 2.7 & 3.7 & 3.3 & \cellcolor{black!10!}3.2 & \cellcolor{black!20!}3.0 \\
    GIN-MLP & 1.7 & 2.0 & 2.7 & \cellcolor{black!10!}2.1 & 1.7 & 2.3 & 3.3 & \cellcolor{black!10!}2.4 & 2.0 & 1.0 & 2.3 & \cellcolor{black!10!}1.8 & \cellcolor{black!20!}2.1 \\
    GIN-\textsc{Trans} & 2.0 & 3.0 & 2.0 & \cellcolor{black!10!}2.3 & 2.0 & 2.7 & 2.3 & \cellcolor{black!10!}2.3 & 1.3 & 2.3 & 1.3 & \cellcolor{black!10!}1.7 & \cellcolor{black!20!}2.1 \\
    \bottomrule
    \end{tabular}
    }
    \end{center}
\end{table}

\begin{figure}
    \centering
    \begin{minipage}{0.48\textwidth}
        \includegraphics[width=\linewidth]{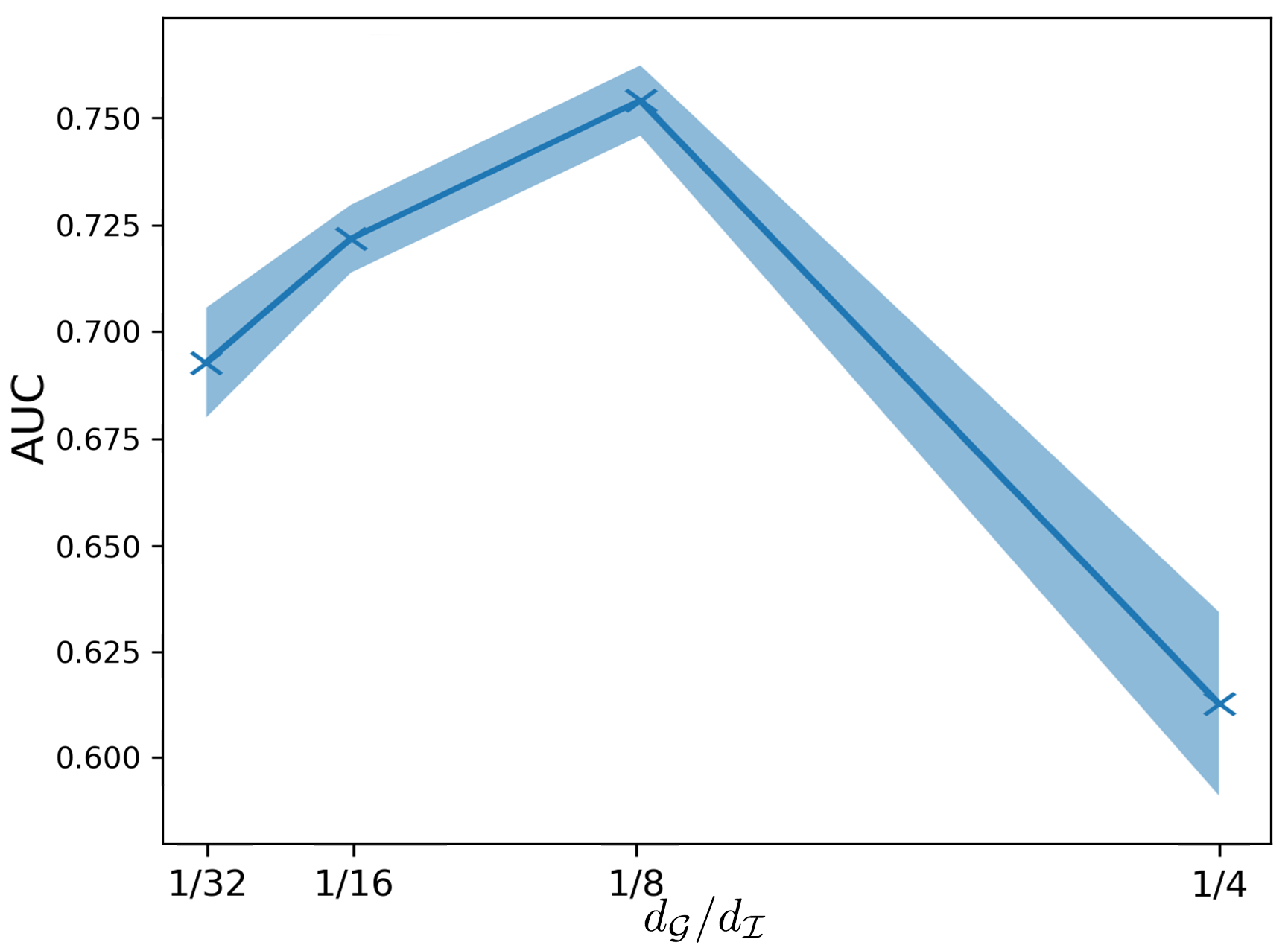}
        \caption{The image-level test AUC of \textbf{CRC-MSI} is sensitive to different choices of $d_\gG/d_\gI$.}
        \label{fig:ablation}
    \end{minipage}
    \hspace{1mm}
    \begin{minipage}{0.48\textwidth}
        \includegraphics[width=\linewidth]{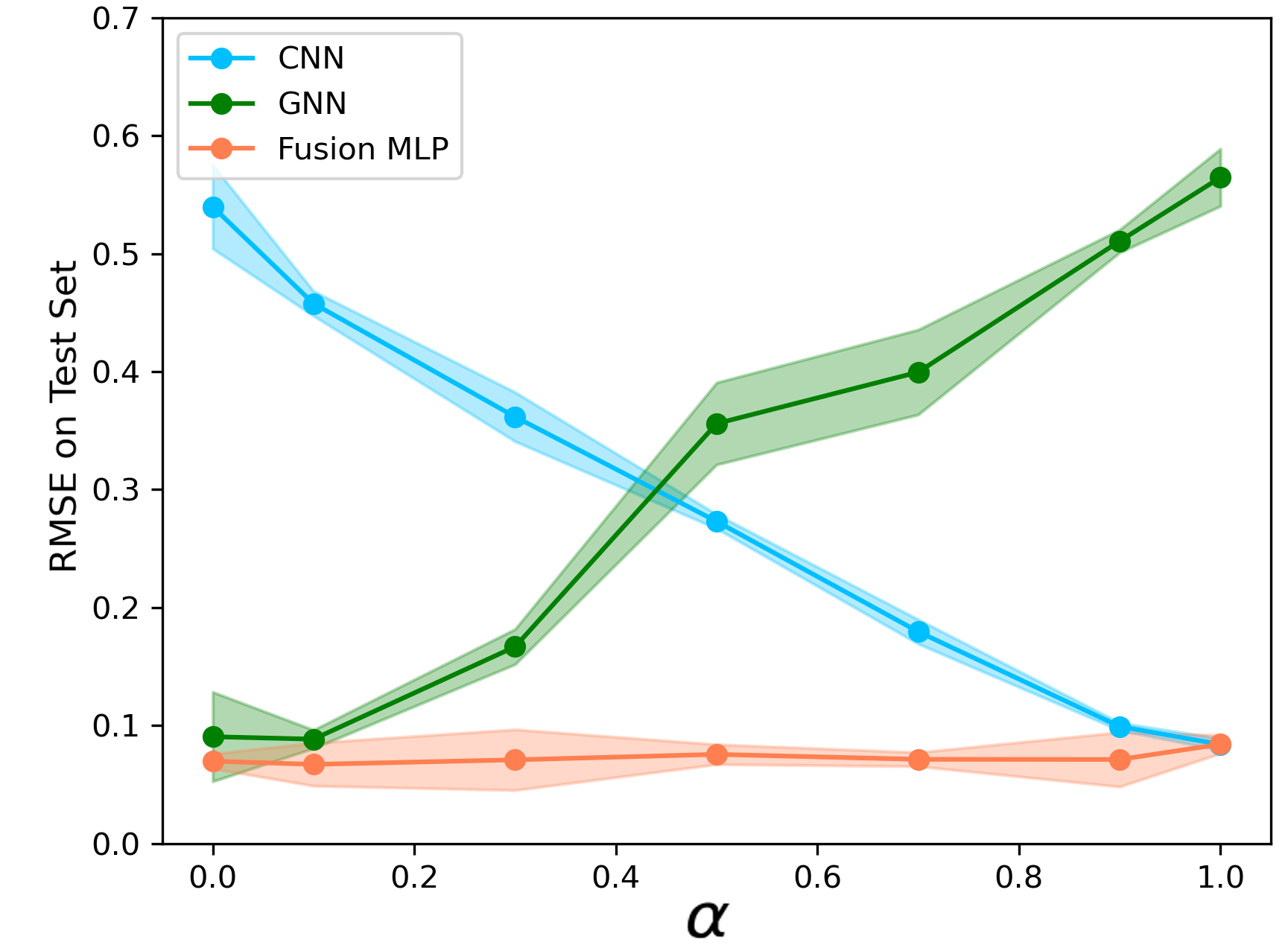}
        \caption{Test RMSE in synthetic task. The fusion model consistently outperforms CNN/GNN.}
        \label{fig:syn_result}
    \end{minipage}
\end{figure}

\subsection{Ablation Study}
In MLP fusion scheme where feature alignment is not necessary, we found the performance is sensitive to the the dimension of $\mH_\gI$, $\mH_\gG$. 
In Figure \ref{fig:ablation}, we present ablations on $d_\gG/d_\gI$ with \textbf{CRC-MSI}, where we use \textsc{GIN} and \textsc{ResNet18} as the backbones and consequently $d_\gI$ is fixed to 512 (see supplementary for details).
Thus, $d_\gG/d_\gI$ needs to be carefully tuned to achieve satisfactory performance in the MLP fusion scheme.
On the contrast, with the feature alignment, \textsc{Transformer} fusion layers are easier to train.

\section{How Fusion Scheme Works?}
In this section, a synthetic task is constructed to explore how the fusion scheme works. 
Specifically, each image $\mX_\gI$ in MNIST and its associated superpixel graph $\mX_\gG$ in MNISTSuperpixel \citep{monti2017geometric} form a paired data $(\mX_\gI,\mX_\gG)$, and we retain their original training and test data partition.
Instead of using the classification label, we synthesize the regression objects which can manually adjust the proportion of image-level and geometric representations:
\begin{equation}
    y = \alpha \cdot f_{\texttt{CNN}}(\mX_\gI) + (1 - \alpha) \cdot f_{\texttt{GNN}}(\mX_\gG),
\end{equation}
where $f_{\texttt{CNN}}$ and $f_{\texttt{GNN}}$ denote the averaged prediction from 30 randomly initialized \textsc{ResNet18} and \textsc{GIN2} (with seed=$0,1,\cdots,29$) respectively (details are elaborated in supplementary). 
The $f_{\texttt{CNN}}(\mathbf{x}_\gI)$ and $f_{\texttt{GNN}}(\mathbf{x}_\gG)$ are fixed once they are generated.
We suppose that $f_{\texttt{CNN}}$ produces objects easier for \textsc{LeNet5} to learn, and $f_{\texttt{GNN}}$ easier for \textsc{GCN2}.
The coefficient $\alpha \in [0,1]$ balances the learning difficulty of CNN or GNN, and also the proportion of image-level and geometric-level features used in model prediction. 
Specifically, the large value of $\alpha$ represents higher proportion of image-level features, thus easier for CNN to learn.
The objective $y$s are normalized for both training and testing.
Comparisons of GNN, CNN and the fusion framework with MLP on $\alpha\in\{0.0,0.1,0.3,0.5,0.7,0.9,1.0\}$ are illustrated in Figure \ref{fig:syn_result}. 
The test RMSE of \textsc{LeNet5} monotonically decrease with a larger $\alpha$; while RMSE of \textsc{GCN} increases.
It yields that \textsc{GCN} fails achieve parallel results as \textsc{LeNet5} when the task is more CNN-friendly (\eg~ $\alpha\ge0.5$ in our setting).
Similarly, when the geometric information is dominant (\eg~ $\alpha\le0.5$), \textsc{LeNet5} fails to capture comprehensive representations. 
Our fusion scheme, performs universally better than either \textsc{LeNet5} or \textsc{GCN} at any ratio between the geometric and morphological representations.
The improvement reaches its peak in the range $\alpha \in [0.3,0.7]$, which coincides with the hypothesized settings in various medical image analysis tasks. 
Furthermore, when the task tends to be more CNN/GNN-friendly, the fusion model, though slight, keeps outperforming CNN/GNN.
It confirms that the features extracted by CNN and GNN are complementary from two views \ie~textural and geometric, thus should be considered jointly.

\section{Related Work}

\paragraph{Deep Learning for Large-scale Medical Imaging and Biomarker Prediction.}
Deep learning has become the mainstream choice to assess disease in gigapixel WSIs \citep{dm2,intro_dl,intro_challenge,intro_dlmia}. 
Various applications are developed upon the success of deep learning and computer vision for histology diagnostic tasks, such as breast cancer segmentation \cite{cruz2014automatic}, prostate cancer detection \citep{litjens2016deep}, and sentinel lymph nodes detection \citep{intro_camelyon}.
Biomarkers \citep{intro_biomarker,intro_why_biomarker} are the clinical indicators for tumor behavior that distinguish patients who will benefit from certain molecular therapies.
Previous works show that deep learning can predict specific genetic mutations in non–small cell lung cancer from hematoxylin and eosin (H\&E) stained slides \citep{intro_biomarker3,intro_biomarker1,intro_biomarker4,intro_biomarker2,intro_biormarker_survey_all}.
Much progress have been made in microsatellite instability (MSI) or mismatch repair deficiency (dMMR) in colorectal cancer (CRC) \citep{intro_msi,intro_msi5,intro_msi6,intro_msi1,intro_msi2,intro_msi3,intro_msi4} Test AUC is risen up to 0.99, which thus even inspired wide commercial interests \citep{intro_msi_review}.

\paragraph{Graph Neural Networks for Medical Imaging.}
CNNs have gained in successfully aiding the diagnostic process for histology arises from their capability to discover hierarchical feature representations without domain expertise \citep{intro_dlmia}. However, the spatial relations, as well as the formulation of structured information, are absent, where this prior is crucial to real clinical diagnosis \citep{barua2018spatial,feichtenbeiner2014critical,galon2006type}. 
GNNs provide an alternative to CNN in its capability of describing relationships and structures \citep{bronstein2017geometric,wu2020comprehensive,zhang2020deep,bronstein2021geometric}.
As a powerful approach to model functional and anatomical structures, the graph-based deep learning approaches have exhibited substantive success for various tasks in the histology domain \citep{lu2021slidegraph+,wang2021cell,tumor_spatial_structure}. 
By capturing the geometrical and topological representations, the mutual interaction between cells can be learned by a neural network.


\section{Discussion}
\label{sec:limitation}
\paragraph{Limitations.}
The fusion scheme takes paired image and graph as input. 
The graph should be extracted prior to the training stage.
However, the construction of the graph is task-specific, with the reliance on domain knowledge. 
Consequently, detailed analysis should be conducted to discover whether the geometric information contributes to the prediction tasks.  
%
%

\paragraph{Broader Impact.}
In the application to clinical practice, fusing the graph representations with image representations enables domain experts to incorporate their prior knowledge to modeling. 
For example, in histology analysis, the construction of cell graphs and the determination of the critical distance reflect how doctors expect cells to interact with each other. 
However, incorporation of the improper prior knowledge to the graph formulations may bring negative effect to the training process, hence should be carefully assessed.
It also points out an interesting future direction towards the automatically generated graph from image with limited human involvement.


\section{Conclusion}
\label{sec:conclusion}
This work proposes a fusion framework for learning the topology-embedded images by CNN and GNN. 
Furthermore, we present a use case of the fusion model to biomarker prediction tasks from histology slides, where the geometric information are proven to be important.
Specifically, we integrate GNN to add local geometric representations for cell-graph patches on top of CNNs which extracts a global image feature representation. 
The CNNs and GNNs are trained in parallel and their output features are integrated in a learnable fusion layer. 
This is important as the fusion scheme addresses the expression of the tumor microenvironment by supplementing topology inside local patches in network training. 
We validate the framework using different combinations of CNN, GNN and fusion modules on real H\&E stained histology datasets, which surpasses the plain CNN or GNN methods to a significant margin.
The experiments yield that the geometric feature and image-level feature are complementary. 
Finally, the constructed image-graph bimodal datasets can serve as benchmark for future study.

\bibliographystyle{ACM-Reference-Format}
\bibliography{reference}

\newpage
\input{appendix}

\end{document}

%% file: math_commands.tex
\usepackage{amsmath,amsfonts,bm}

\def\eqref#1{(\ref{#1})}

\def\1{\bm{1}}

\def\mA{{\bm{A}}}

\def\mH{{\bm{H}}}

\def\mX{{\bm{X}}}

\DeclareMathAlphabet{\mathsfit}{\encodingdefault}{\sfdefault}{m}{sl}
\SetMathAlphabet{\mathsfit}{bold}{\encodingdefault}{\sfdefault}{bx}{n}

\def\gE{{\mathcal{E}}}

\def\gG{{\mathcal{G}}}

\def\gI{{\mathcal{I}}}

\def\gN{{\mathcal{N}}}

\def\gV{{\mathcal{V}}}



%% file: appendix.tex
\appendix
\section{Dataset}
\label{sec:app:data}

This section reveals details for three real-world benchmark datasets. 
We start by reviewing the two public datasets, \ie~\textbf{CRC-MSI} of colorectal cancer patients and \textbf{STAD-MSI} of gastric adenocarcinoma cancer patients for microstatellite status prediction, as well as a privately curated gastric cancer dataset (\textbf{GIST-PDL1}) for PD-L1 status classification. 
In Table~\ref{table:descriptive}, we brief three datasets with the numerical statistics.


\subsection{CRC-MSI and STAD-MSI}
The two public datasets focus on the prediction of distinguishing the microsatellite instability (MSI) from microsatellite stability (MSS) in H\&E stained histology.
Notably, MSI is a crucial clinical indicator for oncology workflow in the determination of whether a cancer patient responds well to immunotherapy. 
It is not until very recently that researches have shown the promising performance of deep learning methods in MSI prediction. 
With the lack of an abundant number of annotated histology, MSI prediction is still very challenging. 
Thus, it is required to incorporate prior knowledge such as geometric representation for MSI prediction.

In the experiment, the two datasets classify images patches to either \textit{MSS} (microsatellite stable) or \textit{MSIMUT} (microsatellite instable or highly mutated). 
We treat \textit{MSIMUT} as the positive label and MSS as the negative label in computing the AUC. 
The original whole-slide images (WSIs) are derived from diagnostic slides with formalin-fixed paraffin-embedded (FFPE) processing. 
In particular, \textbf{CRC-MSI} contains H\&E stained histology slides of $315$ colorectal cancer patients, and \textbf{STAD-MSI} includes H\&E slides of $360$ gastric cancer patients. 
For both datasets, a WSI with respect to a patient is tessellated into non-overlapped patches/images with a resolution of $224\times 224$ pixels at the magnification of 20$\times$. 
The patches from $70\%$ patients are used for training and the remaining patches from $30\%$ patients are left for validation. 
Note that each patient is associated with only one WSI.
Yet, the number of generated image patches from one WSI varies from each other.
Consequently, the ratios of training and test image samples depicted in Table~\ref{table:descriptive} for \textbf{CRC-MSI} and \textbf{STAD-MSI} are not $70\%:30\%$ as its patient-level ratio.

\subsection{GIST-PDL1}
The privately collected \textbf{GIST-PDL1} predicts programmed death-ligand 1 (PD-L1) status from gastric cancer histology slides. PD-L1 is a type of immune-checkpoint protein from tumor cells that disturbs the body's immune system through binding programmed death 1 (PD-1) on T cells. The PD-L1 expression is one of the only established biomarkers that determine the efficacy of immunotherapy in gastric and esophageal cancer in advanced stages \citep{smyth2021checkpoint}. 

This dataset collects $129$ well-annotated H\&E stained histology slides of gastric cancer patients between the year 2020 and the year 2021 from [anonymous] hospital, with ethical approval.
Each whole-slide image (WSI) corresponds to one patient, where the patient is labeled as either \textit{positive} (CPS$\ge5$) or \textit{negative} (CPS$<5$) determined by its PD-L1 combined positive score (CPS) tested from the immunohistochemistry (IHC) test. 
The patch-level annotation inherits the associated patient/WSI-level label.
The resolution of a WSI is around $10,000 \times 10,000$ pixels, which is split into non-overlapping images (patches) of $512 \times 512$ pixels at the magnification $20\times$, and afterward resized to $224\times 224$ to get aligned with the two public datasets. 
Background patches are excluded for downstream analysis with OTSU algorithm, where the remaining patches are subsequently stain normalized to reduce the data heterogeneity. 
Each patch comprises approximately $200$ cells \ie~nodes. 
Different from \textbf{CRC-MSI} and \textbf{STAD-MSI}, we conduct down-sampling on the number of image patches from each WSI to balance the ration between \textit{positive} and \textit{negative} samples. 
Consequently, we achieve a balanced image-level sample ratio that is close to $50\%:50\%$.

\subsection{Data Availability}
\label{sec:app:dataAvailability}
The two public datasets for MSI classification with their annotations freely available at 
$$
\text{\texttt{https://doi.org/10.5281/zenodo.2530834}}.
$$
The private dataset along with the established graph data for three benchmarks will be released after acceptance.

\section{Nuclei Segmentation}
\label{sec:app:seg}
In the node construction process, we employ CA\textsuperscript{2.5}-Net \citep{segment} as the backbone for nuclei segmentation, due to its outstanding performance in challenging clustered edges segmentation tasks, which frequently occurs in histology analysis. 
We use the implementation at \texttt{https://github.com/JH-415/CA2.5-net}.
Specifically, CA\textsuperscript{2.5}-Net formulates nuclei segmentation task in a multi-task learning paradigm that uses edge and cluster edge segmentation to provide extra supervision signals. 
To be more concretely, the decoder in CA\textsuperscript{2.5}-Net comprises three output branches that learn the nuclei semantic segmentation, normal-edge segmentation (\ie~non-clustered edges), and clustered-edge segmentation respectively.
A proportion of the convolutional layers and upsampling layers in the CA\textsuperscript{2.5}-Net is shared to learn common morphological features. 
We follow the original settings \citep{segment} by using the IoU loss for the segmentation path of the nuclei semantic ($\mathcal{L}_{sem}$) and the smooth truncated loss for segmentation paths of normal-edges ($\mathcal{L}_{nor}$) and clustered-edges ($\mathcal{L}_{clu}$). 
Formally, the overall loss thus takes a weighted average over the three terms of segmentation losses, \ie
\begin{equation}
    \mathcal{L} = \alpha \cdot \mathcal{L}_{sem} + \beta \cdot \mathcal{L}_{nor} + \gamma \cdot \mathcal{L}_{clu}.
\end{equation}
In particular, we applies the balancing coefficients $\alpha=0.7$, $\beta=0.2$, and $\gamma=0.1$. 
We trained CA\textsuperscript{2.5}-Net with \textsc{Adam} optimizer for $50$ epochs, with an initial learning rate of $1\times10^{-4}$ that decayed by $0.95$ for every other epochs. 
At the inference stage of nuclei locations, we adopt the nuclei segmentation path to derive the prediction result.

\begin{figure}[th]
    \centering
    \begin{minipage}{0.23\textwidth}
        \includegraphics[width=\linewidth]{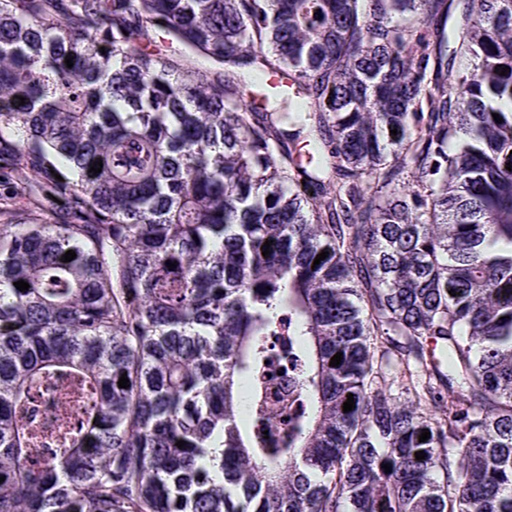}
    \end{minipage}
    \hspace{1mm}
    \begin{minipage}{0.23\textwidth}
        \includegraphics[width=\linewidth]{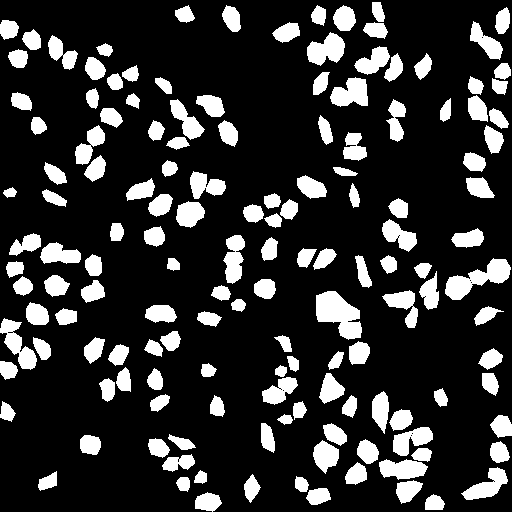}
    \end{minipage}
    \hspace{1mm}
    \begin{minipage}{0.23\textwidth}
        \includegraphics[width=\linewidth]{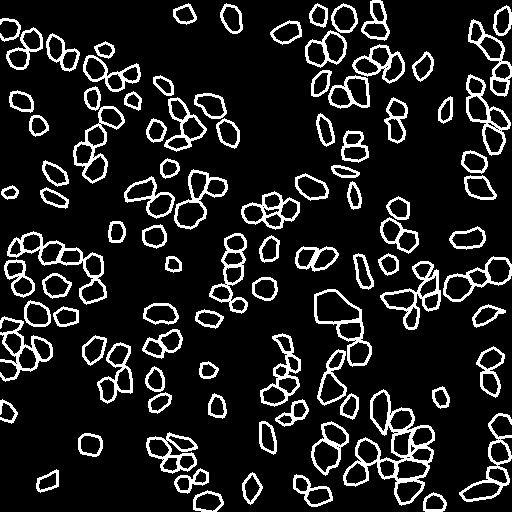}
    \end{minipage}
    \hspace{1mm}
    \begin{minipage}{0.23\textwidth}
        \includegraphics[width=\linewidth]{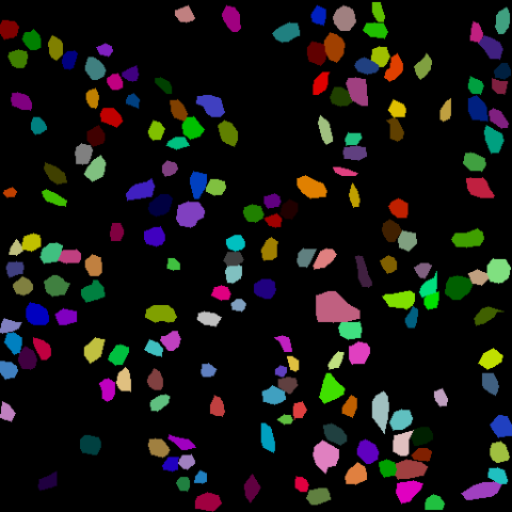}
    \end{minipage}
\caption{An illustrative example of annotated histology patch from \textbf{GIST-PDL1} for training the nuclei segmentation network. The four subgraphs from left to right are the raw patch image, the generated semantic nuclei masks, the generated semantic nuclei edge, and the annotated instance nuclei mask (ground truth).}
\label{fig:nucleiSeg_4_PDL1}
\end{figure}

Three well-experienced pathologists annotated a number of $132$ image patches from \textbf{GIST-PDL1} for training the CA\textsuperscript{2.5}-Net, where we use $100$ images for training and the remaining $32$ for validation. 
To increase the data variations, we adopt offline augmentation (\ie~images are augmented prior to the training stage) by randomly flipping and rotating for 90 degrees.
Eventually, we come to a total number of $400$ training samples.
For illustration purposes, we pick one annotated sample and show it in Figure~\ref{fig:nucleiSeg_4_PDL1}. 
The pixel-level instance annotations were conducted with `\texttt{labelme}' (\texttt{https://github.com/wkentaro/labelme}), where the semantic masks can be generated directly from the instance segmentation \citep{segment}.

\begin{table*}[th]
    \caption{The pre-defined node attributes computed for each nuclei area.}
    \label{tab:attribute}
    \begin{center}
    \resizebox{\linewidth}{!}{
    \begin{tabular}{lll}
    \toprule
    \textbf{GLCM} & \textbf{GLDM} & \textbf{GLRLM} \\
    autocorrelation & dependence entropy & gray-level non-uniformity \\
    cluster  prominence & dependence non-uniformity & gray-level non-uniformity normalized \\
    cluster shade & dependence non-uniformity normalized & gray-level variance \\
    cluster tendency & dependence variance & high gray-level run emphasis \\
    contrast & gray-level non-uniformity & long-run emphasis \\
    correlation & gray-level variance & long-run high gray-level emphasis \\
    difference average & high gray-level emphasis & long-run low gray-level emphasis \\
    difference entropy & large-dependence emphasis & low gray-level run emphasis \\
    difference variance & large-dependence high gray-level emphasis & run entropy \\
    inverse difference & large-dependence low gray-level emphasis & run length non-uniformity \\
    inverse difference moment & low gray-level emphasis & run length non-uniformity normalized \\
    inverse difference moment normalized & small-dependence emphasis & run percentage \\
    inverse difference normalized & small-dependence high gray-level emphasis & run variance \\
    informational measure of correlation 1 & small-dependence low gray-level emphasis & short-run emphasis \\
    informational measure of correlation 2 &  & short-run high gray-level emphasis \\
    Inverse variance & & short-run low gray-level emphasis \\
    joint average & \textbf{FIRST-ORDER} &  \\
    joint energy & 10 percentile & \\
    joint entropy & 90 percentile & \textbf{GLSZM} \\
    maximal	correlation	coefficient & energy & gray-level non-uniformity \\
    maximum probability & entropy & gray-level non-uniformity normalized \\
    sum average & inter quartile range & gray-level variance \\
    sum entropy & kurtosis & high gray-level zone  emphasis \\
    sum squares & maximum & large  area emphasis \\
    & mean absolute deviation & large area high gray-level emphasis \\
    \textbf{LOCATION} & mean & large area low gray-level emphasis \\
    center of mass-x & median & low gray-level zone emphasis \\
    center of mass-y & minimum & size zone non-uniformity \\
     & range & size zone  non-uniformity normalized \\
    \textbf{NGTDM} & robust mean absolute deviation & small area emphasis \\
    busyness & root mean squared & small area high gray-level emphasis \\
    coarseness & skewness & small area low gray-level emphasis \\
    complexity & total energy & zone Entropy \\
    contrast & uniformity & zone Percentage \\
    strength & variance & zone  variance \\
    \bottomrule
    \end{tabular}
    }
    \end{center}
\end{table*}

\section{Node Feature Extraction}
\label{sec:app:tmeAttribute}
This section details the essential pre-processing of the raw histology input (\ie~images) to extract morphological features as node attributes in the construction of cell graphs. 
The same procedure applies to all three datasets. 
The segmentation results of CA\textsuperscript{2.5}-Net on slide patches generate nodes of graphs. 
For an arbitrary patch, a graph is generated where nodes represent cells and the weighted edges reveal the Euclidean distance between nodes. 

Next, we select $94$ features from pathomics, \ie~a pre-defined feature library for medical image analysis \citep{lambin2017radiomics} 
that describe the location, first-order statistics, and the gray-level textural features of each segmented cell. 
To be specific, the five dimensions of the spatial distribution include gray-level co-occurrence (GLCM), gray-level	distance-zone  (GLDM), gray-level run-length (GLRLM), gray-level size-zone (GLSZM), and neighborhood gray tone difference (NGTDM).
In total, there are $2$ coordinates of the cell location, $18$ values of the first-order statistics, $24$ GLCM, $14$ GLDM, $16$ GLRLM, $16$ GLSZM, and $5$ NGTDM. We give the name of all $94$ features in Table~\ref{tab:attribute} for a better understanding. For a detailed calculation of each attribute, we refer interested readers to \cite{lambin2017radiomics}.

\section{Implementation Details}
\label{sec:app:expDetail}
The code is available at: $$\texttt{https://anonymous.4open.science/r/gnncnnfusion2243/}$$
We will replace this anonymous link with a non-anonymous GitHub link after the acceptance. 
All the experiments are implemented in Python 3.8.12 environment on one NVIDIA $^{\circledR}$ Tesla A100 GPU with 6,912 CUDA cores and 80GB HBM2 mounted on an HPC cluster. 
We implement GNNs on PyTorch-Geometric (version 2.0.3) and CNNs on PyTorch (version 1.10.2). 
All CNNs have used \textbf{ImageNet} pre-trained weights provided by TIMM library.

\subsection{Training Settings}
All the model architectures follow the training scheme with the hyper-parameters listed in Table~\ref{tab:searchSpace}. 
We employ the standard cross-entropy as the loss function.
The training stage continues until stopping improvements on the validation set after $8$ consecutive epochs.
\begin{table}[th]
    \caption{Hyper-parameters for training the models.}
    \label{tab:searchSpace}
    \begin{center}
    \begin{tabular}{lr}
    \toprule
    \textbf{Hyper-parameters}  & Value \\
    \midrule
    Initial learning rate &  $5\times10^{-4}$  \\
    Minimum learning rate &  $5\times10^{-6}$ \\
    Scheduler & Cosine Annealing (T\_max=$10$)  \\ 
    Optimizer & AdamW \\ 
    Weight Decay & $1\times10^{-5}$\\
    Num\_workers & $12$ \\
    Batch size & $256$ \\ 
    Maximum epoch number & $100$ \\
    \bottomrule
    \end{tabular}
    \end{center}
\end{table}

\subsection{Model Configuration}
Table \ref{tab:modelConfig} describes the configuration of MLP fusion layer, \textsc{Transformer} fusion layer, \textsc{GCN} and \textsc{GIN} used in this research. The model architectures for all three datasets have adopted the same configuration. 
The output for \textsc{GCN} and \textsc{GIN} are set to the same size 16, without fine-tuning for each dataset, and for CNNs we use the default output size.
In \textsc{Transformer} fusion layer, we use the \textit{Minimization Alignment} strategy to reduce the computational cost.

\begin{table}[th]
    \caption{Default configurations.}
    \label{tab:modelConfig}
    \begin{center}
    \begin{tabular}{l|lr}
    \toprule
    \multirow{4}{*}{MLP} & \# \texttt{MLPBlock} & $1$ \\ 
    & Feature embedding size & $128$ \\
    & Activation & \texttt{Leaky ReLU} \\
    & Dropout rate & $0.1$ \\
    \midrule
    \multirow{6}{*}{\textsc{Transformer}} & \# \texttt{TransBlock} &  $1$ \\
    & Feature embedding size & $192$ \\
    & Activation & \texttt{ReGLU} \\
    & \# Attention heads & $4$ \\
    & Dropout rate in \texttt{TransBlock} & $0.1$ \\
    & Alignment strategy & \textit{Minimization Alignment} \\
    \midrule
    \multirow{3}{*}{\textsc{GCN} and \textsc{GIN}} & \# layers & $2$ \\ 
    & Feature embedding size & $128$ \\
    & Activation & \texttt{GeLU} \\
    \bottomrule
    \end{tabular}
    \end{center}
\end{table}

\subsection{Data Augmentations}
To improve the model generalization, we perform online data augmentation for both graphs and images in the training stage. 
The graph data is augmented with RandomTranslate where the translate is set to 5. 
Images are first resized to $224\times 224$ pixels, and then augmented by Random Horizontal Flip, Random Vertical Flip, and Random Stain Normalization and Augmentation. 
Random Stain Normalization and Augmentation is designed to spec targets to alleviate the stain variations specifically for histology, where stain normalization template is randomly generated per iteration.

\subsection{Comparison on the Number of Trainable Parameters}
Table~\ref{tab:paramsNum} reports the number of trainable parameters of all the models we evaluated in Table~\ref{tab:classification}. The values are given with the image input size of $3\times224\times224$ and the node feature dimension of $94$. The scale of the trained model is jointly determined by the choice of modules in CNN, GNN, and fusion layers, where we highlighted different options by color. The choice of colors aligns with the associated modules visualized in Figure~\ref{fig:architecture_model}. For instance, green colors include three selections of CNN modules, including \textsc{MobileNetV3}, \textsc{DenseNet}, and \textsc{ResNet}. `N/A' indicates an absence of such layers in the framework. The numbers are reported in millions ($1\times10^6$). For instance, $13.1277$ at the bottom-right of the table means that it involves $13.1277$ millions of learnable parameters when training an integrated model with the \textsc{ResNet18}-\textsc{GIN2}-\textsc{Trans} architecture. 

\begin{table}[th]
    \caption{Comparison on the number of trainable parameters (in millions).}
    \label{tab:paramsNum}\vspace{-3mm}
    \begin{center}
    \begin{tabular}{llrrrr}
    \toprule
    & & \multicolumn{4}{c}{\color{green!60!black!80!}\textbf{CNN}}\\ \cmidrule(lr){3-6} 
    \color{orange!90!white!70!}\textbf{GNN} & \color{blue!60!cyan!70!black!67!}\textbf{Fusion} & \cellcolor{green!60!black!80!}N/A & \cellcolor{green!60!black!80!}\textsc{MobileNetV3} & \cellcolor{green!60!black!80!}\textsc{DenseNet} & \cellcolor{green!60!black!80!}\textsc{ResNet} \\
    \midrule
    \cellcolor{orange!90!white!70!}N/A     
    & \cellcolor{blue!60!cyan!70!black!67!}N/A & - &    $1.7865$    & $7.2225$ & $11.3110$ \\
    \midrule
    \cellcolor{orange!90!white!70!}
    & \cellcolor{blue!60!cyan!70!black!67!}MLP &   0.0665   &    $1.8506$   & $7.2866$ & $11.3751$ \\
    \multirow{-2}{*}{\cellcolor{orange!90!white!70!}GCN} & \cellcolor{blue!60!cyan!70!black!67!}\textsc{Transformer} &  -   &     $8.4943$  & $13.9303$ & $13.1231$ \\
    \midrule
    \cellcolor{orange!90!white!70!}
    & \cellcolor{blue!60!cyan!70!black!67!}MLP &  $0.0712$   &   $1.8552$    & $7.2912$ & $11.3797$ \\
    \multirow{-2}{*}{\cellcolor{orange!90!white!70!}GIN} & \cellcolor{blue!60!cyan!70!black!67!}\textsc{Transformer} &   -  &  $8.4989$     & $13.9349$ & $13.1277$ \\
    \bottomrule
    \end{tabular}
    \end{center}
\end{table}

\section{Further Performance Analysis }
\label{sec:app:analysis}
\begin{table*}[htbp]
    \caption{Test Acc and AUC improvements of the fusion model to pure CNN on three benchmarks.}
    \vspace{-1mm}
    \label{tab:classification_improvement}
    \begin{center}
    \resizebox{\linewidth}{!}{
    \begin{tabular}{llrrrrrrrrr}
    \toprule
    && \multicolumn{3}{c}{\textbf{GIST-PDL1}} & \multicolumn{3}{c}{\textbf{CRC-MSI}} & \multicolumn{3}{c}{\textbf{STAD-MSI}}\\ \cmidrule(lr){3-5}\cmidrule(lr){6-8}\cmidrule(lr){9-11}
    \multicolumn{2}{c}{\textbf{Model}} & $\Delta$ACC & $\Delta$AUC & $\Delta$AUC\textit{\footnotesize{patient}} & $\Delta$ACC & $\Delta$AUC & $\Delta$AUC\textit{\footnotesize{patient}} & $\Delta$ACC & $\Delta$AUC & $\Delta$AUC\textit{\footnotesize{patient}} \\
    \midrule
    \multirow{4}{*}{\rotatebox[origin=c]{90}{\footnotesize{MobileNetV3}}} 
    & \textsc{GCN}-\textsc{MLP} & $3.56$ & $1.96$ & $8.06$ & $-0.25$ & $7.17$ & $12.87$ & $0.78$ & $4.98$ & $0.35$\\
    & \textsc{GIN}-\textsc{MLP} & $1.33$ & $2.59$ & $10.32$ & $0.43$ & $2.79$ & $13.47$ & $0.53$ & $2.90$ & $0.46$ \\
    & \textsc{GCN}-\textsc{Trans} & $4.27$ & $3.69$ & $12.82$ & $0.19$ & $4.75$ & $13.16$ &  $1.01$ & $6.73$ & $1.53$ \\
    & \textsc{GIN}-\textsc{Trans} & $2.56$ & $4.16$ & $10.51$ & $0.51$ & $4.24$ & $14.89$ & $0.95$ & $6.47$ & $1.65$\\
    \midrule
    \multirow{4}{*}{\rotatebox[origin=c]{90}{DenseNet}} 
    & \textsc{GCN}-\textsc{MLP} & $5.35$ & $6.35$ & $6.89$ & $1.04$ & $0.07$ & $1.60$ & $1.66$ & $8.96$ & $1.04$\\
    & \textsc{GIN}-\textsc{MLP} & $5.58$ & $5.51$ & $5.61$ & $0.46$ & $0.16$ & $4.95$ & $2.07$ & $9.25$ & $0.58$\\
    & \textsc{GCN}-\textsc{Trans} & $8.45$ & $7.29$ & $8.47$ & $0.81$ & $4.01$ & $16.60$ & $1.77$ & $8.04$ & $1.44$ \\
    & \textsc{GIN}-\textsc{Trans} & $4.64$ & $5.20$ & $6.99$ & $0.65$ & $4.64$ & $9.08$ & $1.90$ & $8.82$ & $0.98$\\
    %
    \midrule
    \multirow{4}{*}{\rotatebox[origin=c]{90}{ResNet}} 
    & \textsc{GCN}-\textsc{MLP} & $11.25$ & $10.50$ & $7.81$ & $0.62$ & $9.84$ & $21.50$ & $2.26$ & $0.50$ & $0.66$ \\
    & \textsc{GIN}-\textsc{MLP} & $5.61$ & $5.51$ & $5.04$ & $0.99$ & $4.54$ & $21.33$ & $2.43$ & $2.13$ & $1.49$ \\
    & \textsc{GCN}-\textsc{Trans} & $5.39$ & $4.31$ & $7.39$ & $1.26$ & $8.33$ & $23.04$ & $2.35$ & $0.10$ & $1.43$\\
    & \textsc{GIN}-\textsc{Trans} & $6.16$ & $9.99$ & $9.27$ & $1.08$ & $8.12$ & $22.13$ & $2.42$ & $1.99$ & $1.71$\\
    \bottomrule\\[-2.5mm]
    \end{tabular}
    }
    \vspace{-3mm}
    \end{center}
\end{table*}

Table~\ref{tab:classification_improvement} below reveals the absolute percentage improvement of each metrics in the main prediction tasks. The comparisons are made on basis of CNNs methods. To be specific, an absolute improvement score is calculated by 
\begin{equation*}
    \Delta\text{ Score}=S_{\text{fused}}-S_{\text{CNN}},
\end{equation*}
where $S_{\text{CNN}}$ denotes the performance score (\ie~ ACC, AUC or AUC\textit{\footnotesize{patient}}) achieved by plain CNNs (\eg~\textsc{MobileNetV3}, \textsc{DenseNet} or \textsc{ResNet}), and $S_{\text{fused}}$ is the associated score by integrated models, which are listed in the first column at the very left. For instance, the $3.56$ at the top-left of the table means that the test accuracy of \textsc{MobileNetV3}-\textsc{GCN}-\textsc{MLP} is improved by $3.56\%$ to \textsc{MobileNetV3}.

\section{Visualization of Nuclei Segmentation and Cell Graph}
\begin{figure}[htbp]
    \begin{minipage}{0.23\textwidth}
        \includegraphics[width=\linewidth]{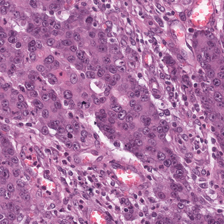}
    \end{minipage}
    \hspace{1mm}
    \begin{minipage}{0.23\textwidth}
        \includegraphics[width=\linewidth]{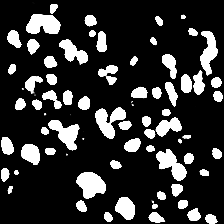}
    \end{minipage}
    \hspace{1mm}
    \begin{minipage}{0.23\textwidth}
        \includegraphics[width=\linewidth]{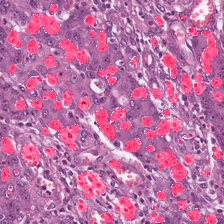}
    \end{minipage}
    \hspace{1mm}
    \begin{minipage}{0.23\textwidth}
        \includegraphics[width=\linewidth]{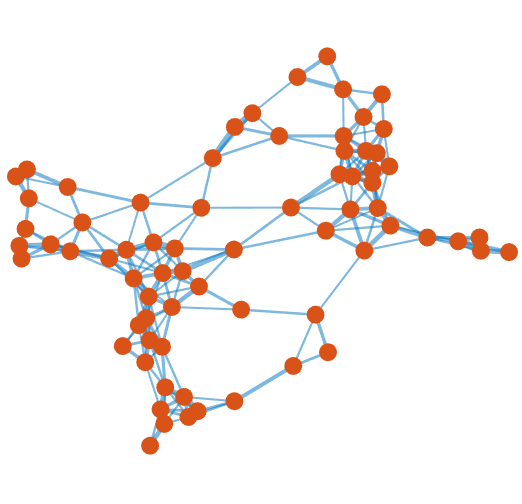}
    \end{minipage}
    \caption{Visualization of the segmented cells and the generated graphs from an arbitrary patch sample of \textbf{CRC-MSI}. The four subgraphs from left to right are the raw patch image, the segmented cells masks, the patch image with overlaid segmentation masks, and the generated graph. }
    \label{fig:nucleiSeg_graph_CRC}
\vspace{5mm}
    \begin{minipage}{0.23\textwidth}
        \includegraphics[width=\linewidth]{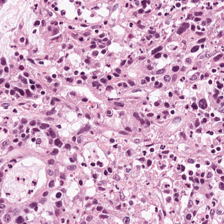}
    \end{minipage}
    \hspace{1mm}
    \begin{minipage}{0.23\textwidth}
        \includegraphics[width=\linewidth]{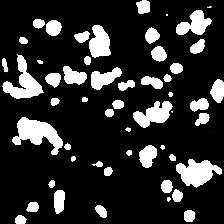}
    \end{minipage}
    \hspace{1mm}
    \begin{minipage}{0.23\textwidth}
        \includegraphics[width=\linewidth]{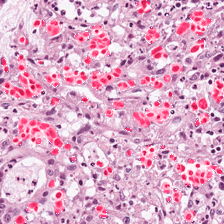}
    \end{minipage}
    \hspace{1mm}
    \begin{minipage}{0.23\textwidth}
        \includegraphics[width=\linewidth]{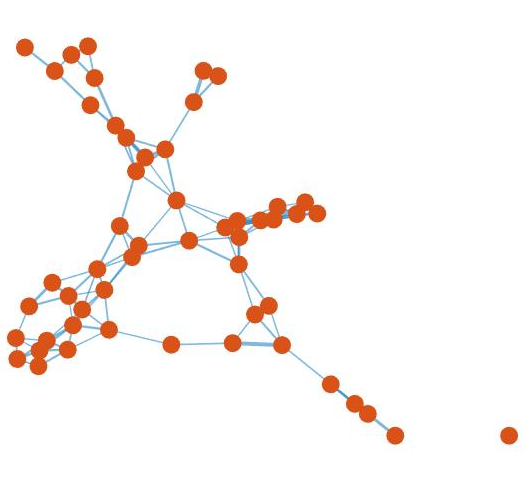}
    \end{minipage}
    \caption{Visualization of the segmented cells and the generated graphs from an arbitrary patch sample of \textbf{STAD-MSI}. The four subgraphs from left to right are the raw patch image, the segmented cells masks, the patch image with overlaid segmentation masks, and the generated graph.}
    \label{fig:nucleiSeg_graph_STAD}
\end{figure}

To better understand the learned graphs that are generated from histology images, Figures~\ref{fig:nucleiSeg_graph_PDL1}-\ref{fig:nucleiSeg_graph_STAD} investigate some random patch images from the three datasets and visualize the nuclei segmentation results and the associated graphs. In particular, the four subgraphs from left to right of each figure display the raw patch image, the segmented cells masks, the patch image with overlaid segmentation masks, and the generated graph.

\section{Synthetic Task}

\paragraph{Data.} 
The MNIST images are provided by \texttt{torchvision.datasets.mnist}, and the associated superpixel graph is provided by \texttt{torch\_geometric.datasets.mnist\_superpixels}. 
Images and graphs are paired based on their sample index. 

\begin{figure}[htbp]

\begin{minipage}{0.48\textwidth}
        \includegraphics[width=\linewidth]{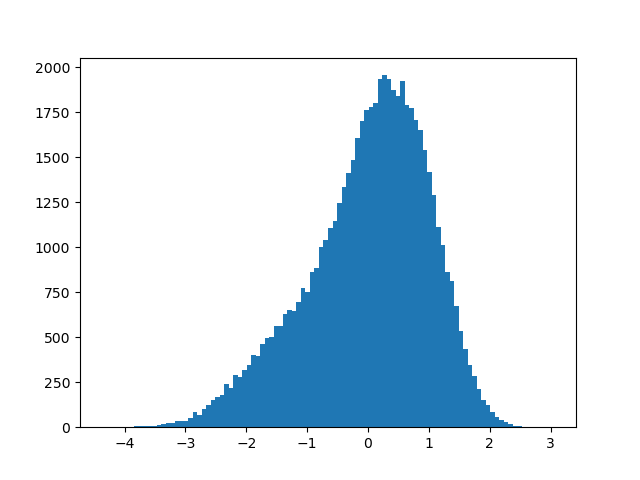}
    \end{minipage}
    \hspace{1mm}
\begin{minipage}{0.48\textwidth}
        \includegraphics[width=\linewidth]{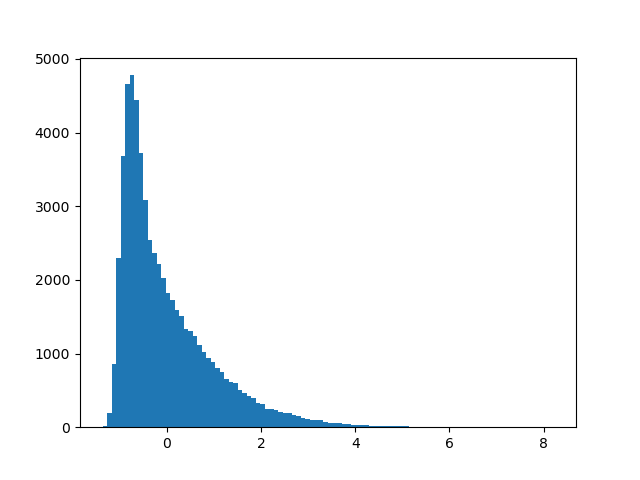}
    \end{minipage}
    \hspace{1mm}
    \caption{The distribution of normalized $f_{\texttt{CNN}}(\mX_\gI)$ (left) and $f_{\texttt{GNN}}(\mX_\gG)$ (right) on the training set.
    }\label{fig:app_syn_label_dis}
\end{figure}

\paragraph{Regression Label Normalization.} We normalize $f_{\texttt{CNN}}(\mX_\gI)$ separated for train and test sets by: 
\begin{equation}
    f_{\texttt{CNN}}(\mX_\gI)_\text{normalized} = \frac{
        f_{\texttt{CNN}}(\mX_\gI)_\text{raw} -
        \texttt{Mean}
        \big(f_{\texttt{CNN}}(\mX_\gI)_\text{raw}\big)
    }{
    \texttt{STD}\big(f_{\texttt{CNN}}(\mX_\gI)_\text{raw}\big)
    },
\end{equation}
where we use the subscript `normalized' and `raw' to denote the post-processed and pre-processed targets.
$f_{\texttt{GNN}}(\mX_\gG)$ is normalized in the same fashion. 
The histogram in Figure \ref{fig:app_syn_label_dis} depicts the label distribution on the training set.

\begin{figure}[htbp]

\begin{minipage}{0.48\textwidth}
        \includegraphics[width=\linewidth]{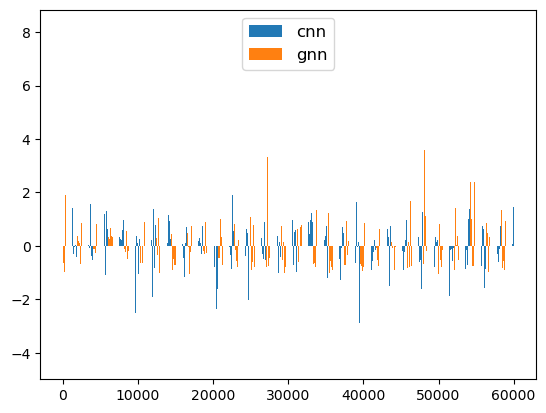}
    \end{minipage}
    \hspace{1mm}
\begin{minipage}{0.48\textwidth}
        \includegraphics[width=\linewidth]{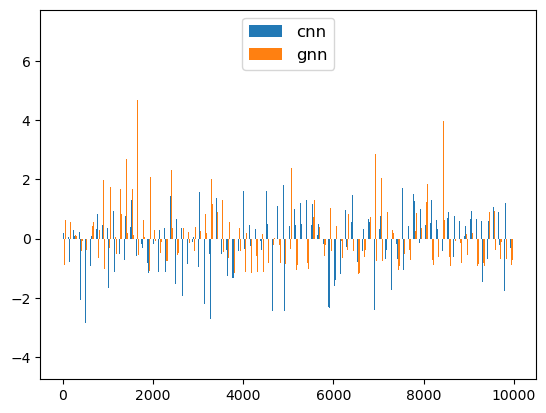}
    \end{minipage}
    \hspace{1mm}
    \caption{The normalized $f_{\texttt{CNN}}(\mX_\gI)$ and $f_{\texttt{GNN}}(\mX_\gG)$ used to construct synthetic label $y$ for training set (left) and test set (right).
    The $x$-axis is the sample index provided by the original datasets.
    }\label{fig:app_syn_label}
\end{figure}

\paragraph{Label Distributions.} In Figure \ref{fig:app_syn_label}, a significant difference can be observed from the $f_{\texttt{CNN}}(\mX_\gI)$ and $f_{\texttt{GNN}}(\mX_\gG)$, which matches our hypothesis that they are easy for different structures to learn. 

\paragraph{Model Configurations.} The configurations of fusion MLP and \textsc{GIN} for synthetic tasks are depicted in Table \ref{tab:modelConfig2}. A clear difference can be observed between the ground truth of $f_{\texttt{CNN}}(\mX_\gI)$ and $f_{\texttt{GNN}}(\mX_\gG)$. The \textsc{LeNet5} follows the default configurations.

\begin{table}[htbp]
    \caption{Hyper-parameter configurations for the synthetic task.}
    \label{tab:modelConfig2}
    \begin{center}
    \begin{tabular}{l|lr}
    \toprule
    \multirow{4}{*}{MLP} & \# \texttt{MLPBlock} & $1$ \\ 
    & Feature embedding size & $10$ \\
    & Activation & \texttt{Leaky ReLU} \\
    & Dropout rate & $0.1$ \\
    \midrule
    \multirow{3}{*}{\textsc{GIN} } & \# layers & $2$ \\ 
    & Feature embedding size & $32$ \\
    & Activation & \texttt{GeLU} \\
    \bottomrule
    \end{tabular}
    \end{center}
\end{table}

\paragraph{Training Scheme.} We select mean square error (MSE) as the loss function, with other hyper-parameters presented in Table \ref{tab:searchSpace2}. We use root mean square error (RMSE) as the evaluation metric. We stop training if the test RSME does not decrease for 10 epochs. For each $\alpha$, we perform 5 random runs and report the mean and standard deviation. 

\begin{table}[htbp]
    \caption{Hyper-parameters for training the model in the synthetic task.}
    \label{tab:searchSpace2}
    \begin{center}
    \begin{tabular}{lr}
    \toprule
    \textbf{Hyper-parameters}  & Value \\
    \midrule
    Scheduler & Constant \\ 
    Optimizer & \texttt{AdamW} \\ 
    Learning rate &  $1\times10^{-4}$  \\
    Weight Decay & $1\times10^{-5}$\\
    Num\_workers & $10$ \\
    Batch size & $256$ \\ 
    Maximum epoch number & $10,000$ \\
    \bottomrule
    \end{tabular}
    \end{center}
\end{table}

\begin{figure*}[htbp!]
    \centering
    \includegraphics[width=0.55\linewidth]{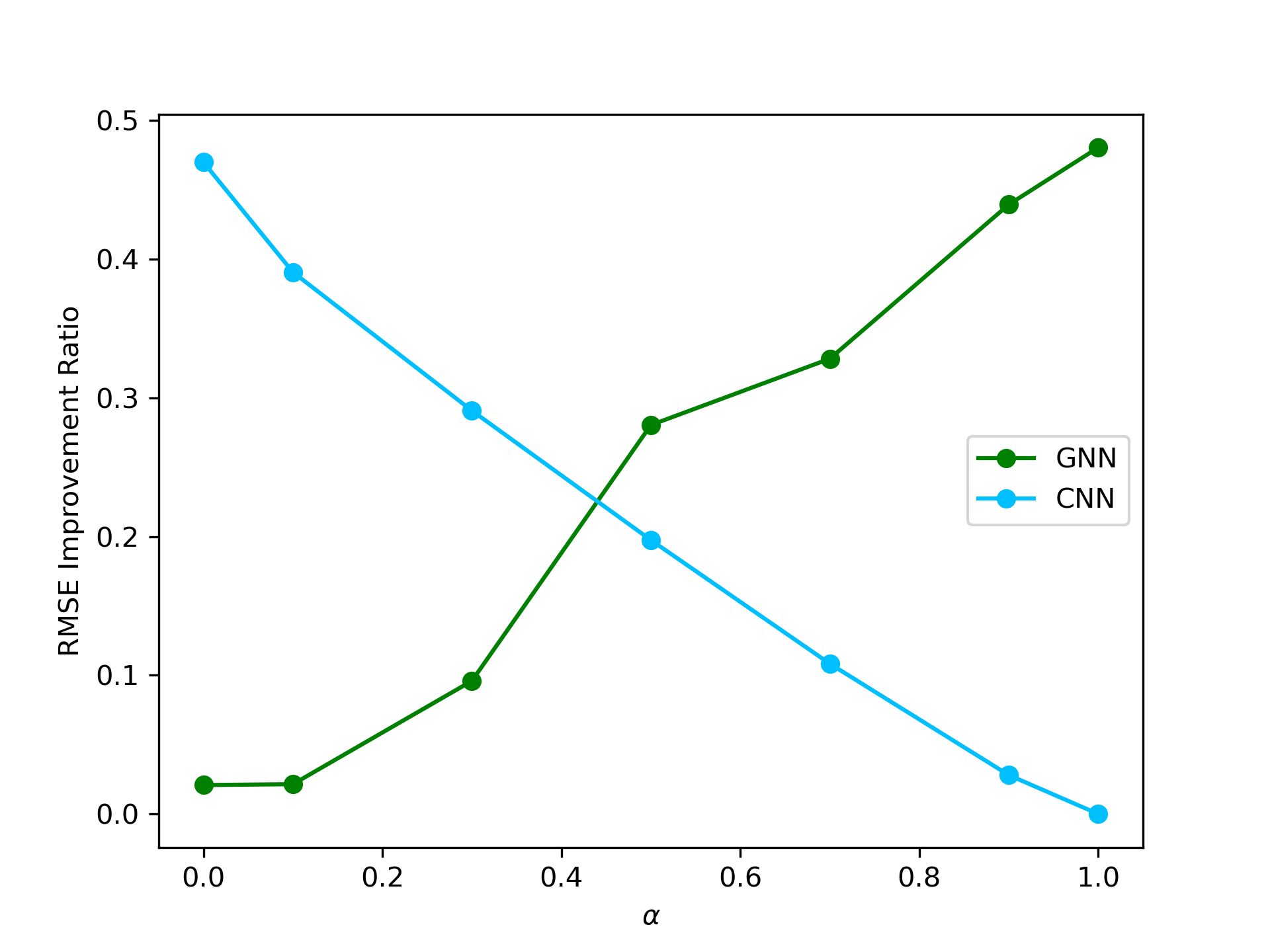}
    \vspace{-3mm}
    \caption{
    RMSE improvement to to \textsc{LeNet5} and \textsc{GIN2} on the test set.
    }
    \vspace{-3mm}
    \label{fig:app_syn_improve}
\end{figure*}

\paragraph{Analysis.} The performance improvement of fusion model to CNN (\textsc{LeNet5}) and GNN (\textsc{GIN2}) is shown in Figure \ref{fig:app_syn_improve}.

\section{Ablation Study}
\paragraph{Number of the \texttt{MLPBlocks}.} For MLP fusion scheme, we use \textsc{GIN} and \textsc{ResNet18} on \textbf{CRC-MSI} to explore the effect of number of \texttt{MLPBlocks} (\# \texttt{MLPBlocks}). The image-level AUCs are $75.2\pm0.85$, $73.4\pm1.74$, $73.1\pm0.82$ for \# \texttt{MLPBlocks}=1, 2, 3. Consequently, more MLP layers does not contribute to better performance.

\section{Related Works}
\paragraph{Background for Pathology.}
Histology analysis has received much attention, because they are widely considered as the gold standard for cancer diagnosis \citep{intro_pathology}.
The complex patterns and tissue structures presented in histology make manual examination very labor-intensive and time-consuming \citep{intro_pathology2,intro_pathology3}.
Even well-experienced experts take a long time to perform careful manual assessments for one slide by observing the histological section under microscopes.
With the advent of computational pathology techniques, glass slides are scanned into high-resolution Whole Slide Images (WSIs) for computer-assisted interventions. 
The remarkable information density of digital histology empowers the complete transition from feature engineering to the usage of deep learning (DL) in mining extensive data. 
Various applications are developed upon deep learning for histology diagnostic tasks, such as breast cancer segmentation \citep{cruz2014automatic}, prostate cancer detection \citep{litjens2016deep}, sentinel lymph nodes detection \citep{intro_camelyon}.

\paragraph{Background for Biomarker Prediction from Histology.}
Biomarkers are defined as \textit{clinical indicators for tumor behavior \eg~the responsiveness to therapy or recurrence risk} \citep{intro_biomarker}.
Identification of biomarkers helps distinguish patients who can benefit from certain molecular therapies \citep{intro_why_biomarker}.
However, with an increasing amount of biomarkers in oncology workflow, the complexity of treatment recommendations increases tremendously \citep{intro_biomarker_difficulty2}. 
Furthermore, both the molecular and immunotherapy approaches for biomarker recognition require either additional tissue material or expensive staining dyes, making it cost-intensive and time-consuming.
As a ubiquitous image source in real clinical practice that is routinely prepared for cancer diagnosis, histology provides substantive information to be mined. 
The gene mutations associated with the biomarker rewrite the cellular machinery and change their behavior. 
Although these morphological changes are subtle to be noticed, empirical results yield that deep learning can directly and reliably detect these features from histology slides.
Previous works have shown the effectiveness of deep learning in genetic mutation prediction from non–small cell lung cancer (NSCLC) H\&E slides, such as serine/threonine kinase 11 (STK11), tumor protein p53 (TP53), epidermal growth factor receptor (EGFR) \citep{intro_biomarker1}.
Similar to NSCLC, the expression of programmed death-ligand 1 (PD-L1) can be predicted by a multi‑field‑of‑view neural network model \citep{intro_biomarker2}.
In prostate cancer, an ensemble of deep neural networks can predict the mutation of BTB/POZ protein (SPOP), achieving a test AUC up to 0.74 \citep{intro_biomarker3}.
Another study in melanoma yields that deep learning can predict the NRAS proto-oncogene (NRAS) and B-Raf proto-oncogene (BRAF) from H\&E slides \citep{intro_biomarker4}. 
In terms of microsatellite instability (MSI) or mismatch repair deficiency (dMMR) in colorectal cancer (CRC), following the initial proof of the concept published in 2019 \citep{intro_msi}, a line of research papers \citep{intro_msi1,intro_msi2,intro_msi3,intro_msi4,intro_msi5,intro_msi6} have made much progress. 